\documentclass[aps,prl,amsmath,amssymb,superscriptaddress,notitlepage]{revtex4-1} 
\usepackage[utf8]{inputenc}
\usepackage[english]{babel}
\usepackage[titletoc,toc,title]{appendix}
\usepackage{amsmath}
\usepackage{amsfonts}
\usepackage{amssymb}
\usepackage[colorlinks=true,linkcolor=blue,citecolor=blue,urlcolor=blue]{hyperref}
\usepackage{graphicx}
\usepackage{enumerate}
\usepackage{csquotes}
\usepackage[caption=false]{subfig}
\usepackage{dsfont}
\usepackage{color}
\usepackage{bbold}
\usepackage{mathtools}
\usepackage{tensor}
\usepackage[export]{adjustbox}
\usepackage{hyperref}

\usepackage{epigraph}
\begin{document}

\title{Classical Hydrodynamics for Analogue Spacetimes:\\ Open Channel Flows and Thin Films.}

\author{Germain Rousseaux}
\affiliation{Institut Pprime, CNRS--Universit\'{e} de Poitiers--ISAE ENSMA, Futuroscope, France}

\author{Hamid Kellay}
\affiliation{Universit\'e de Bordeaux, LOMA, Talence, France.}


\begin{abstract}
Here we review the way to build analogue spacetimes in open channel flows by looking at the flow phase diagram and the corresponding analogue experiments performed during the last years in the associated flow regimes. Thin films like the circular jump with different dispersive properties are discussed with the introduction of a brand new system for the next generation of analogue gravity experiments: flowing soap films with their capillary/elastic waves.
\end{abstract}

\maketitle

\textit{Once upon a time, an astronomer had a plumbing problem in his house but (s)he did not have time to settle it since (s)he was looking at pictures of the aether (or covariant quantum vacuum) taken with his/her telescope. (S)he called a plumber.  P: Hello, may I help you? A: The bathtub drain is blocked: the water can't evacuate. And the astronomer gets back to his/her telescope images. But the plumber likes to talk when (s)he is working: P: What are you studying?  A: Black holes, wormholes, white holes... and you? P: Drain holes, piping tubes, water taps... The plumber used a suction cup to unplug the debris accumulated in the canalization. P: Look! The dirty things are now going inside the draining vortex. For sure, they will never come back... A: Thanks. P: May I use your bathroom sink to wash my hands? A: Sure! By the way, the tap is leaking. The plumber stopped the water supply outside the house and then fixed the tap before opening again the water entry. Some air was trapped in the process. P: Look! The air bubbles are expulsed from the circular jump. For sure, they won't climb up... A: What is found in the depths of the maelstrom??? Where is going all this water?  P: Ah... It is a lllooonnnngggg journey: one does not know exactly... one should maybe plunged your telescope into the drain hole? A: No way! Gosh, the price is expensive as usual for a plumber but you saved my life! Moral of the story: astronomers should discuss with plumbers...}

\section{Introduction}

According to Thibault Damour, {\it Stephen Hawking is a sphinx who leaves us with enigmas}. Shortly after his death, many commentators wondered why he did not get the Nobel prize: it is said that his most mind-blowing prediction, particles creation by astrophysical black holes, surely will not be observed because of the size of the effect. In 1974, Stephen Hawking, willing to show that black holes were not thermodynamical objects, predicted to his own surprise that classical black holes radiate quantum mechanically \cite{Hawking74} demonstrating Jacob Bekenstein's guess that the area of a black hole behaves like an entropy \cite{Bekenstein73}. He confirmed the concept of black hole temperature, which is proportional to the surface gravity (the tidal acceleration) at the black hole horizon, the place where the speed of light matches the escape velocity hindering the interior of the black hole from the rest of the universe. The theoretical prediction of the Bekenstein-Hawking entropy is considered as a test bed for the yet-to-be-defined theory of Quantum Gravity. Unfortunately, the temperature is so small with respect to the Cosmic Microwave Background glow (the original light from the Big Bang) that it is as if one wanted to observe a firefly in a car headlight or to hear a whisper in a rock concert: hence, there is little chance of observing Hawking radiation in an astrophysics context despite the recent direct observation of the accretion disk of a black hole with the Event Horizon Telescope \cite{Akiyama19}.

\begin{figure}[!h]
\includegraphics[width=2.5in]{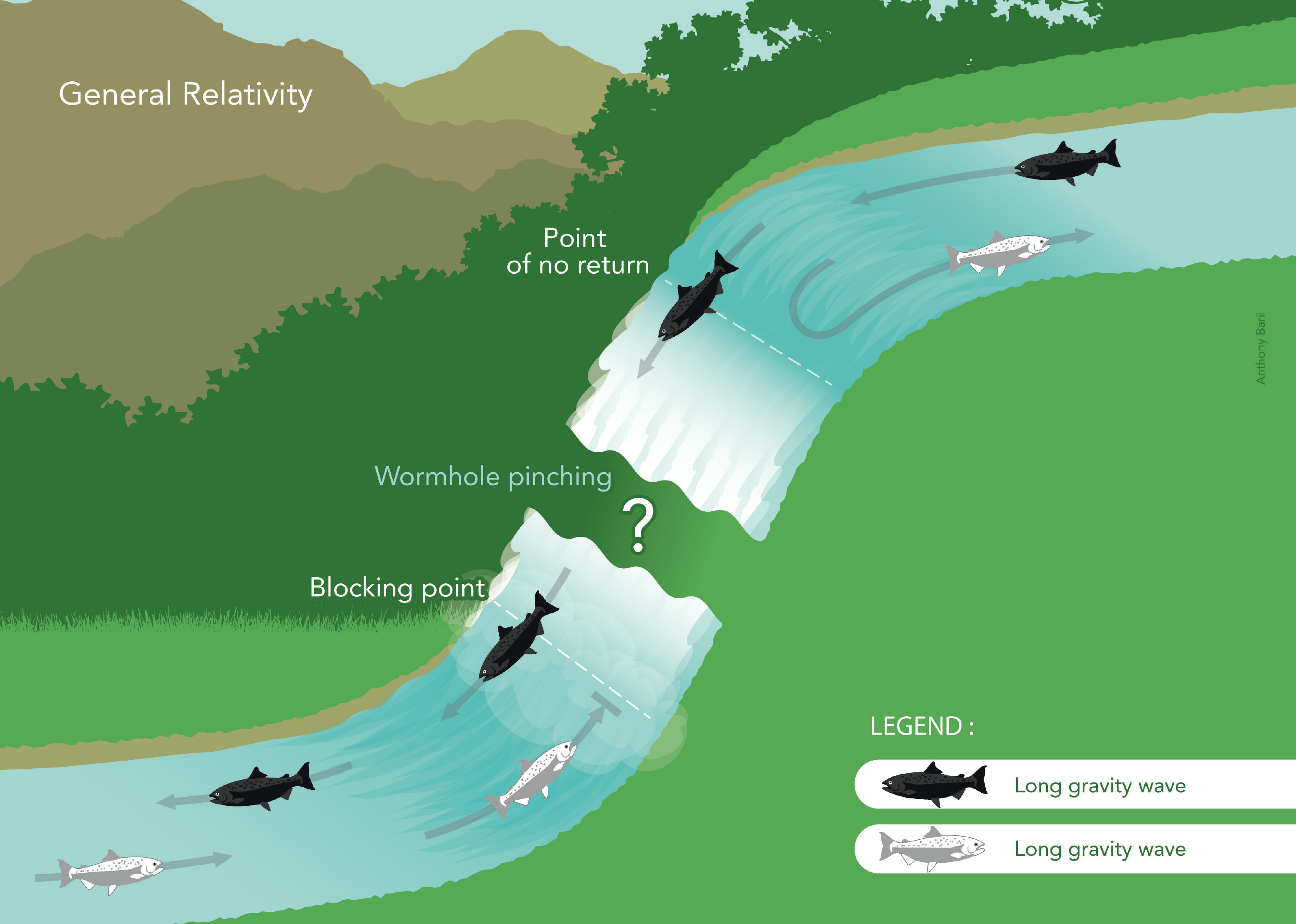}
\includegraphics[width=2.5in]{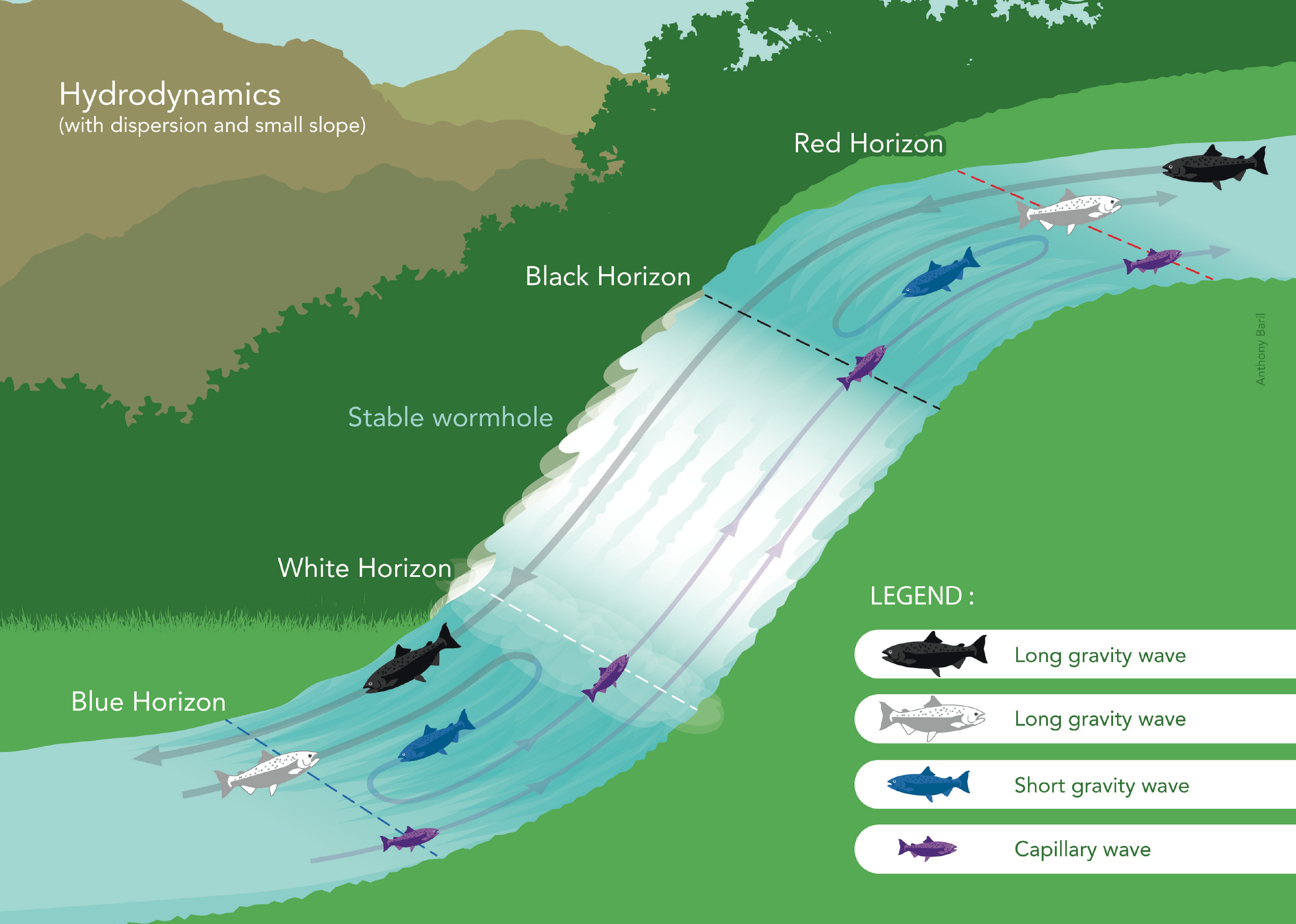}
\includegraphics[width=2.5in]{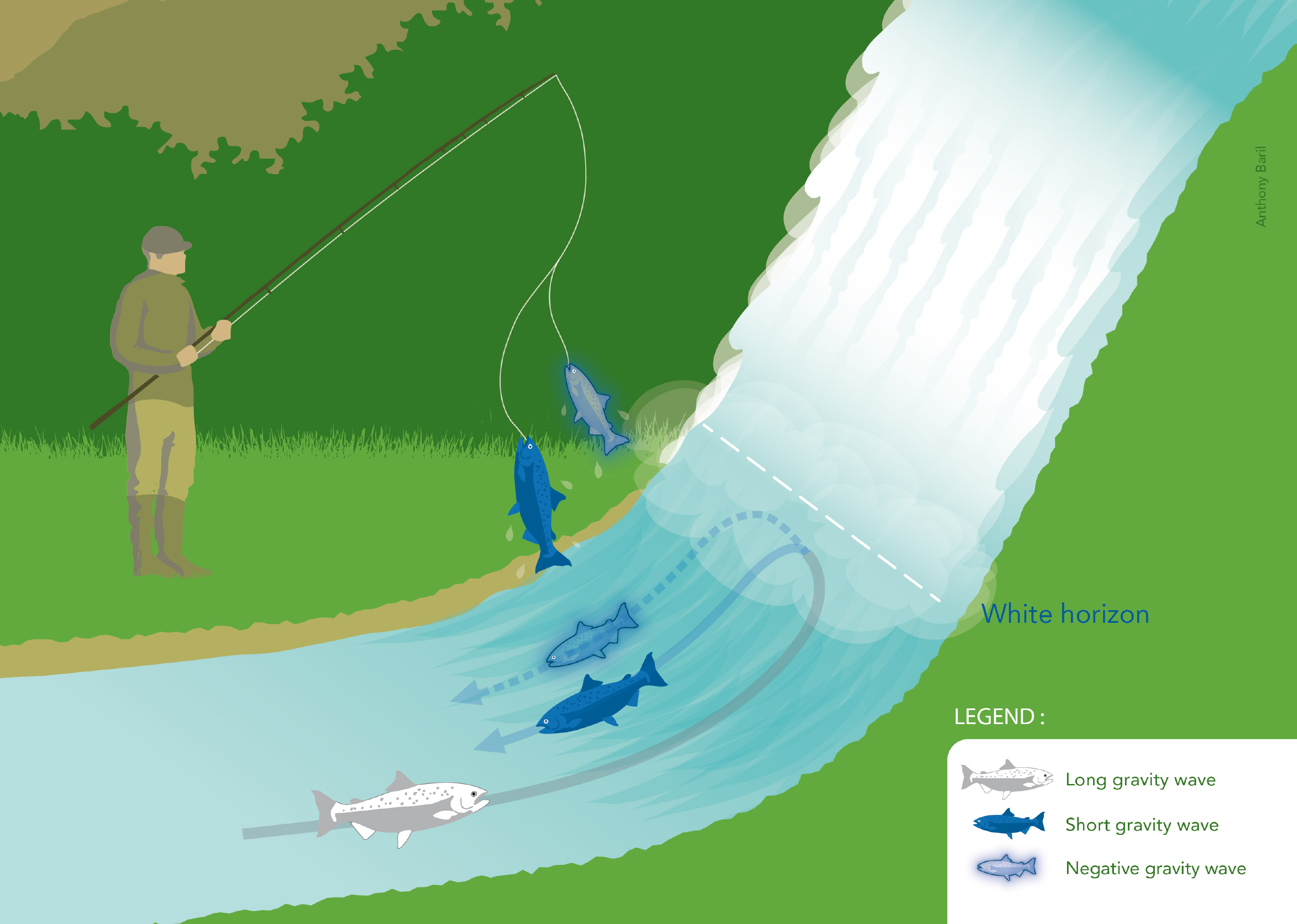}
\includegraphics[width=2.5in]{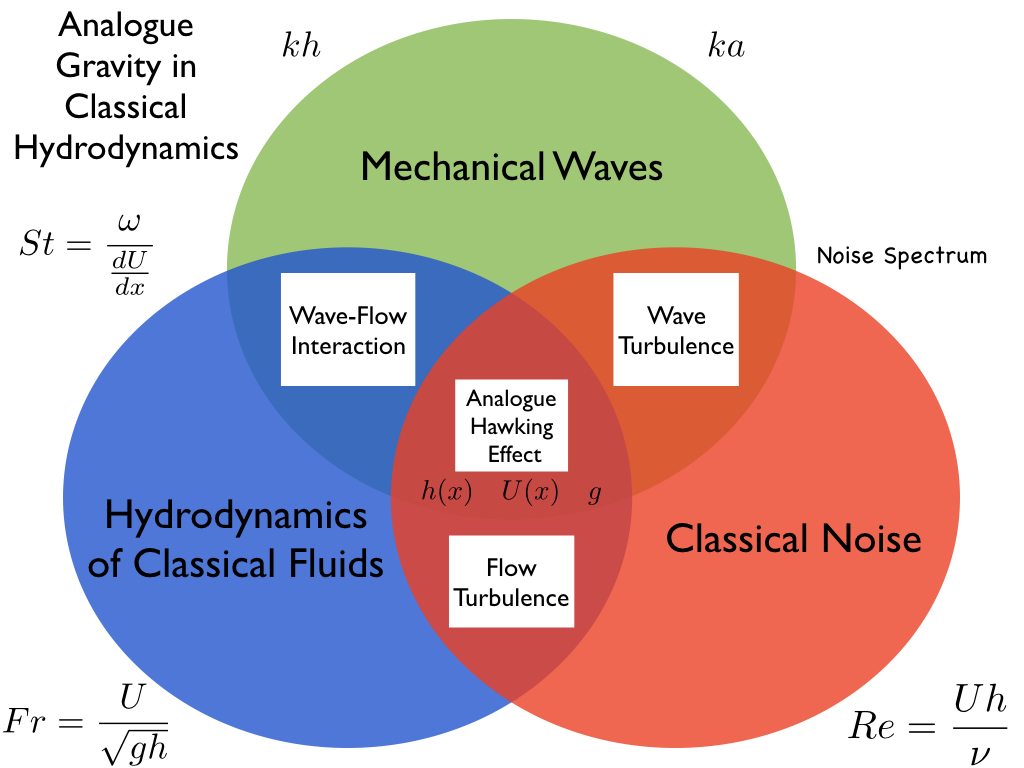}
\caption{(Top) The space-time river analogy: the analogue of the event horizon is the "point of no return" for a swimming fish ; the water flows faster on the left than a fish may swim. If a fish happens to swim beyond this frontier, it can never go back upstream; at the end of the fall, the fish may splash on a smooth body of water or crash on stones, a singular fate akin to the central singularity of astrophysical black holes, or the fish may exit in a decelerating region. By time reversing these statements, the waterfall or drain (the black hole) becomes a fountain or cataract (a white hole) and the region in-between is analogous to a wormhole. In General Relativity, the wormhole is unstable and nobody knows what is inside the singularities of black and white holes. In Hydrodynamics, the flows are usually smoother with no singularity and stable ; When including dispersion, multiple mode conversions happen. (Bottom) The fisherman analogy: (dispersive) mode conversions and the analogue Hawking effect can be characterized by the tool of correlations \cite{Barcelo11, Barcelo19} at equal time akin to a simultaneous fishing by a rod with a double hook ; The flower petals summarize the tryptych met in any condensed matter system amenable to an analogue behaviour: the flowing space-time, the probing waves and the natural noise.}
\label{pictures}
\end{figure}

Nevertheless, Hawking radiation may be observed in the laboratory using analogue space-times \cite{Barcelo11, Como11, Robertson12, Barcelo19} with, for example, light in fibres with varying refractive index \cite{Rosenberg20, Bermudez20}, optical field fluctuations in self-defocusing media \cite{Marino08}, fluid of light in polariton microcavity \cite{Jacquet20}, sound in an inhomogeneous wind \cite{White73, Unruh81}, water waves on an inhomogeneous water current in 1D \cite{Schutzhold02, Como11} and in 2D \cite{Torres20}, phonons in a flowing Bose-Einstein condensate \cite{Garay00}, ripplons in superfluid Helium \cite{Volovik03}, spin waves on an Ohmic current \cite{Jannes11b, Duine17}, etc... And the last years have been characterized by a wealth of experimental achievements in all these areas of condensed-matter, atomic and optical Physics \cite{Barcelo11, Robertson12, Barcelo19}. Indeed, classical and superfluid hydrodynamics flows mimick space-time metrics with horizons. They are relying on an analogy discovered by William Unruh in 1981 for acoustic waves and in 2002 for water waves \cite{Unruh81, Schutzhold02}: the black hole is like a river flowing towards a waterfall, with a potential singularity like the one at the centre of a real black hole. Imagine that the river carries waves of constant velocity (see Figures \ref{pictures}). Suppose that the closer the river gets to the waterfall the faster it flows and that at some point the speed of the river exceeds the waves velocity. The water waves can no longer propagate upstream beyond this zone: the point of no return is similar to a gravitational horizon. The space-time geometry of the gravitational field behaves like an effective flow \cite{Barcelo11, Robertson12, Barcelo19}. The analogue Hawking temperature is proportional to the normal acceleration of the medium as it crosses the horizon which plays the role of the surface gravity at the event horizon in General Relativity \cite{Unruh81, Visser98}. 

Let us dismiss from the start many misconceptions about Hawking radiation: NO, this is not a pure quantum effect that can only occur in astrophysics! It turns out that Hawking prediction was made historically in an astrophysical context with gravitational objects like black holes which act as amplifiers (like a Hi-fi system) of quantum noise (as discussed by Unruh \cite{Unruh11, Unruh14}) namely pairs of particles-antiparticles that would be separated by the tidal forces in the vicinity of an event horizon \cite{Hawking74}. Hence, the Hawking effect is primarily a wave mechanical effect, with many purely classical attributes that can be verified in a classical analogue gravity experiment. Its astrophysics exemplification has a classical counterpart in the mode mixing between positive and negative energy modes that will be detailed in the following as modes of opposite relative frequency in moving fluid media: the flow gradient scattering shows amplification of classical waves leading to spontaneous emission when the waves are quantized. The question of the thermality of the emission depends on the dispersive character or not of the system as discussed by Philbin and the extensive references therein \cite{Philbin16}: in the dispersionless limit, there is a thermal spectrum of Hawking radiation with a gain factor of the negative modes given by the Planckian law when dissipation or greybody factors can be neglected. Back to the gravitational case, the falling anti-particles have negative energies/masses which induce the evaporation of the black hole, a dynamical effect described by Einstein's equations of General Relativity contrary to the pairs separation which is a pure kinematical process described by wave mechanics and which does not rely on the dynamics of gravity. Hence, many moving media featuring perturbations propagating on the top of them can be mathematically characterized by an effective metric with an analogue horizon, the place where the velocity of the medium matches the waves speed provided we are in the so-called hydrodynamic regime, namely a long wavelength approximation without dispersion \cite{Barcelo11, Barcelo19}. The derivation of Hawking is plagued with a caveat namely the so-called trans-Planckian problem \cite{Unruh95}: the wavelength of the outgoing modes goes to zero at the horizon below the Planck scale where one enters possibly into a Quantum Gravity regime where General Relativity would break down as a theory... 

\begin{figure}[!h]
\includegraphics[width=4.5in]{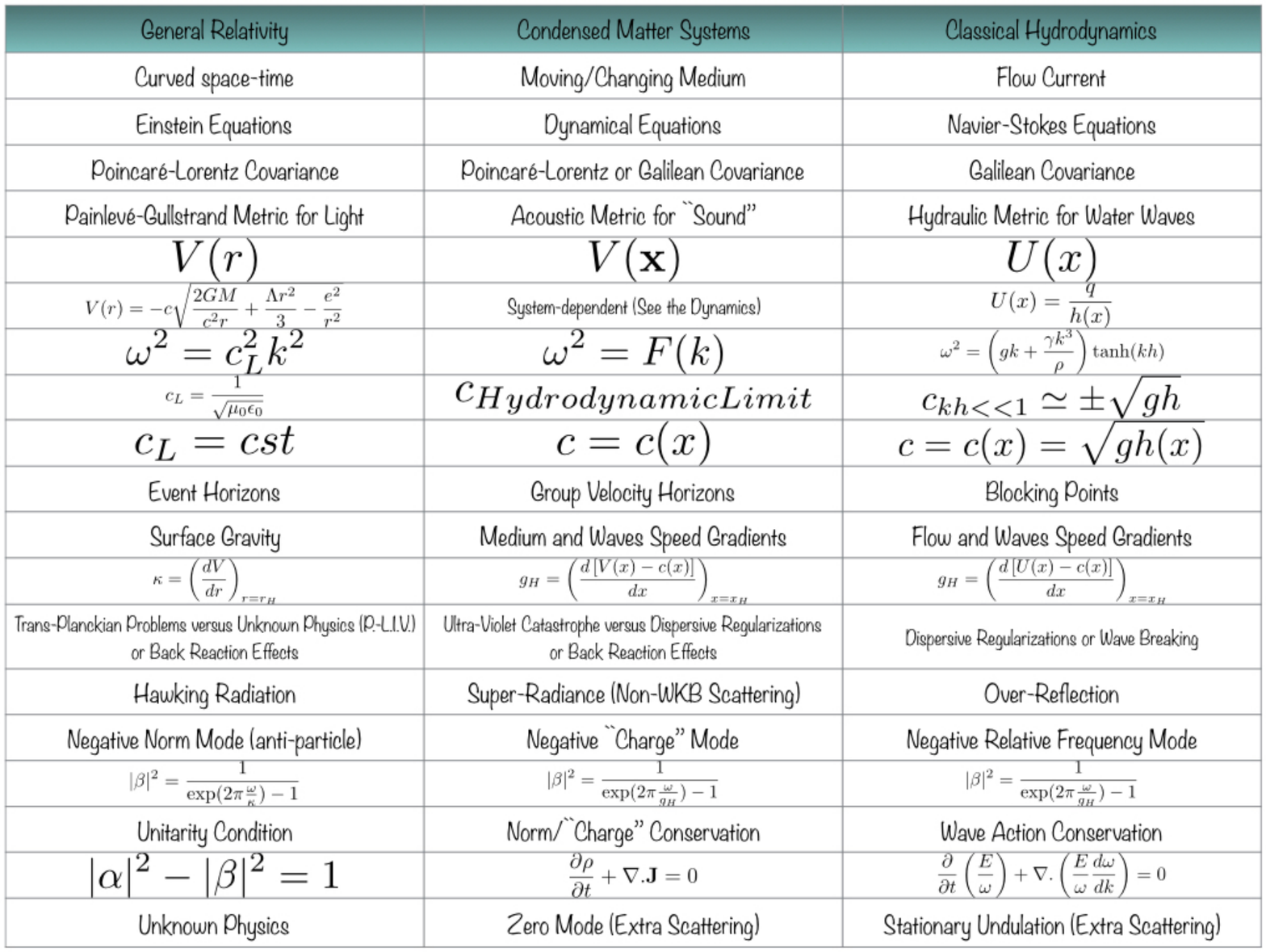}
\caption{The Rosetta Stone of Analogue Gravity establishing the matching between General Relativity, Condensed Matter Systems and Classical Hydrodynamics \cite{Schutzhold02, Rousseaux08, Unruh08, Nardin09, Rousseaux10, Weinfurtner11, Jannes11a, Jannes11b, Como11, Lawrence12, Vazquez12, Faltot13, Coutant14, Faltot14, Euve14, Michel14, Peloquin15, Rojas15, Euve15, Euve16a, Peloquin16, Robertson16, Coutant16, Euve16b, Coello17, Churilov17a, Churilov17b, Euve17, EuvePhD17, Michel18a, Robertson18, Coutant18, Michel18b, Fourdrinoy18, Soto19, Euve20}.}
\label{rosetta}
\end{figure}

Hawking radiation is an anomalous scattering process, which takes place in effective or real space-time geometries with a given metric and applies equally to either noise (be it classical or quantum) or stimulated waves. In the long-wavelength limit, both the system in general relativity and its analogues in condensed matter physics are described by a(n) (effective) metric: $ds^2=c^2dt^2-(dx-Udt)^2$ in 1D for simplicity \cite{White73, Unruh81}. $U$ is the speed of the flow, $c$ is the speed of the waves. Strictly speaking, the kinematic analogy would be broken when dispersive effects are taken into account (by "entering the white hole" and/or "escaping from the black hole") since the Lorentz invariance would not be satisfied anymore as well as the mapping to an effective metric. Both flow velocity $U(x)$ and wave speed $c(x)$ profiles are tuned experimentally. $c$ is usually constant in General Relativity (except in some black hole solutions of modified gravity theories \cite{Martel01}) but may vary in analogue systems. We summarize the mathematical analogy in a table (the "Rosetta Stone") in the Figure \ref{rosetta}.

As far as Classical Hydrodynamics is concerned, the seminal work by Sch\"{u}tzhold \& Unruh in 2002 \cite{Schutzhold02} triggered the first generation of experiments and theoretical developments in 2006-2011 \cite{Rousseaux08, Unruh08, Nardin09, Rousseaux10, Weinfurtner11, Jannes11a, Jannes11b, Como11} that culminate with the SIGRAV Como Summer School in 2011 then a second generation with reproductions, improvements and further comprehension during the years 2012-2019 \cite{Lawrence12,Vazquez12, Faltot13, Coutant14, Faltot14, Euve14, Michel14, Peloquin15, Rojas15, Euve15, Euve16a, Peloquin16, Robertson16, Coutant16, Euve16b, Coello17, Churilov17a, Churilov17b, Euve17, EuvePhD17, Michel18a, Robertson18, Coutant18, Michel18b, Fourdrinoy18, Soto19, Euve20}. The Royal Society meeting on the next generation of analogue gravity experiments in December 2019 was the occasion to assess the present state of the field and possible routes to the future. The present paper stems from the latter.

\section{Analogue Gravity in Open Channel Flows (or the plumber expertise)}

\epigraph{{\it ``By a physical analogy I mean that partial similarity between the laws of one science and those of another which makes each of them illustrate the other. . . . [W]e find the same resemblance in mathematical form between two different phenomena.''}}{--- \textup{James Clerk Maxwell}, \textit{On Faraday's Lines of Force} (1855).}

We present here a short review on open channel flows for Analogue Gravity \cite{Schutzhold02, Rousseaux08, Unruh08, Nardin09, Rousseaux10, Weinfurtner11, Jannes11a, Jannes11b, Como11, Lawrence12, Vazquez12, Faltot13, Coutant14, Faltot14, Euve14, Michel14, Peloquin15, Rojas15, Euve15, Euve16a, Peloquin16, Robertson16, Coutant16, Euve16b, Coello17, Churilov17a, Churilov17b, Euve17, EuvePhD17, Michel18a, Robertson18, Coutant18, Michel18b, Fourdrinoy18, Soto19, Euve20} and some guidelines for future works. When transposed in the Fourier space amenable to a wave analysis, the acoustic metric formalism resumes to the study of the dispersion-less dispersion relation for propagating waves in a moving medium. For instance, if a wave is generated with an angular frequency $\omega$ in presence of a uniform motion of the medium with speed $U$, the "acoustic" dispersion relation in presence of motion is $(\omega -{\bf U.k})^2=c^2k^2$ where $k=\pm \sqrt{k_x^2+k_y^2}$ is the wave number and $c$ is the wave speed (the phase and group speeds are identical in the hydrodynamic limit of long wavelengths with respect to any dispersive scale in the system) \cite{Schutzhold02, Como11}. When dispersive corrections are not taken into account, the spectrum of long gravity water waves is described by the effective dispersion relation using relativistic notation $g^{\mu\nu}k_\mu k_\nu =0$ with the so-called effective metric tensor in contravariant form $g^{\mu\nu}$ and wavevector in covariant form $k_\mu=(\omega, -k_x, -k_y)$. The contravariant components of the metric are given by $g^{00}=-1$, $g^{0i}=-{\bf U}^i$ and $g^{ij}=c^2\delta ^{ij}-{\bf U}^i{\bf U}^j$ \cite{Schutzhold02, Como11, Lawrence12}.

\begin{figure}[!h]
\includegraphics[width=2.6in]{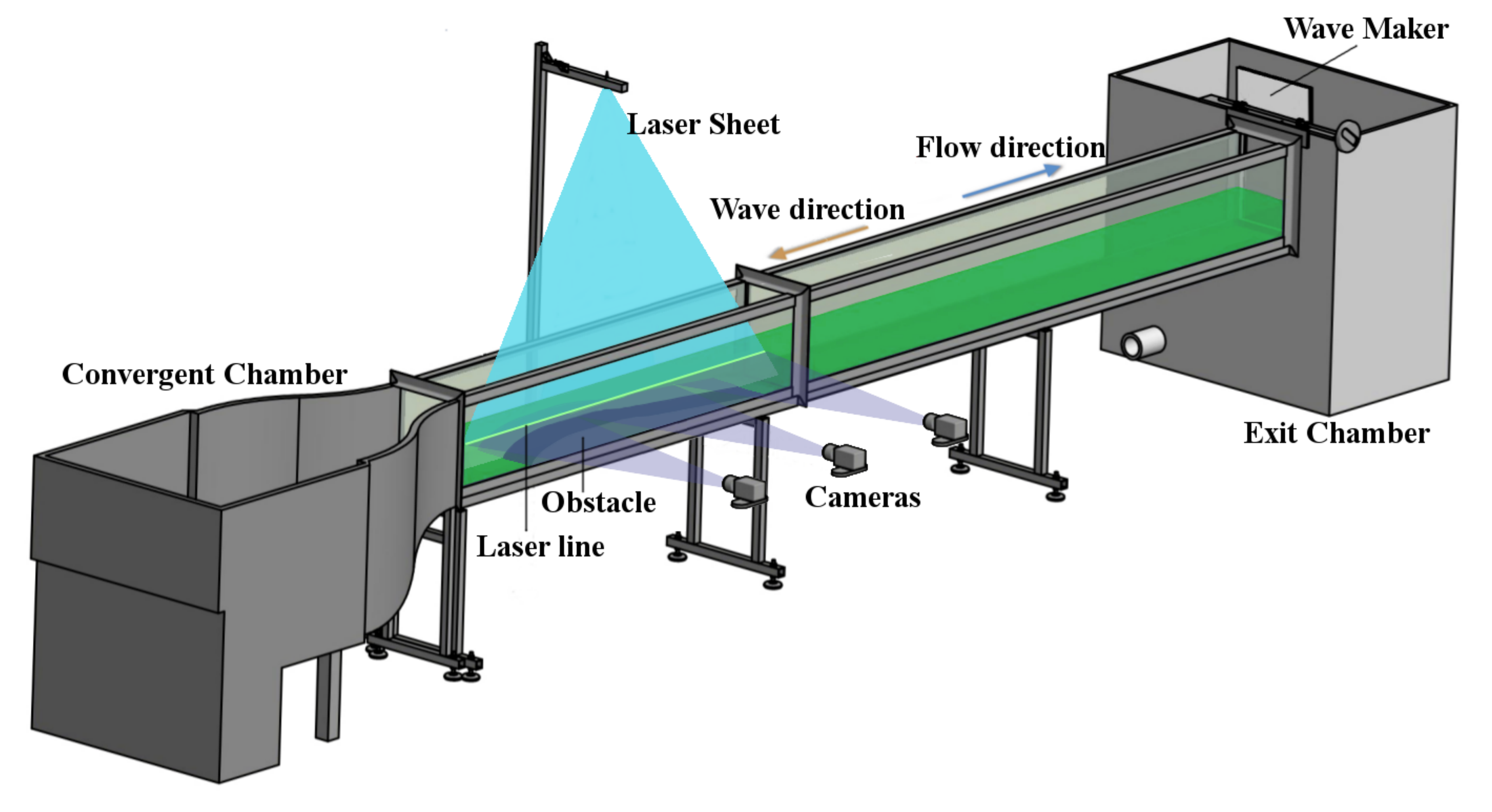}
\includegraphics[width=2.6in]{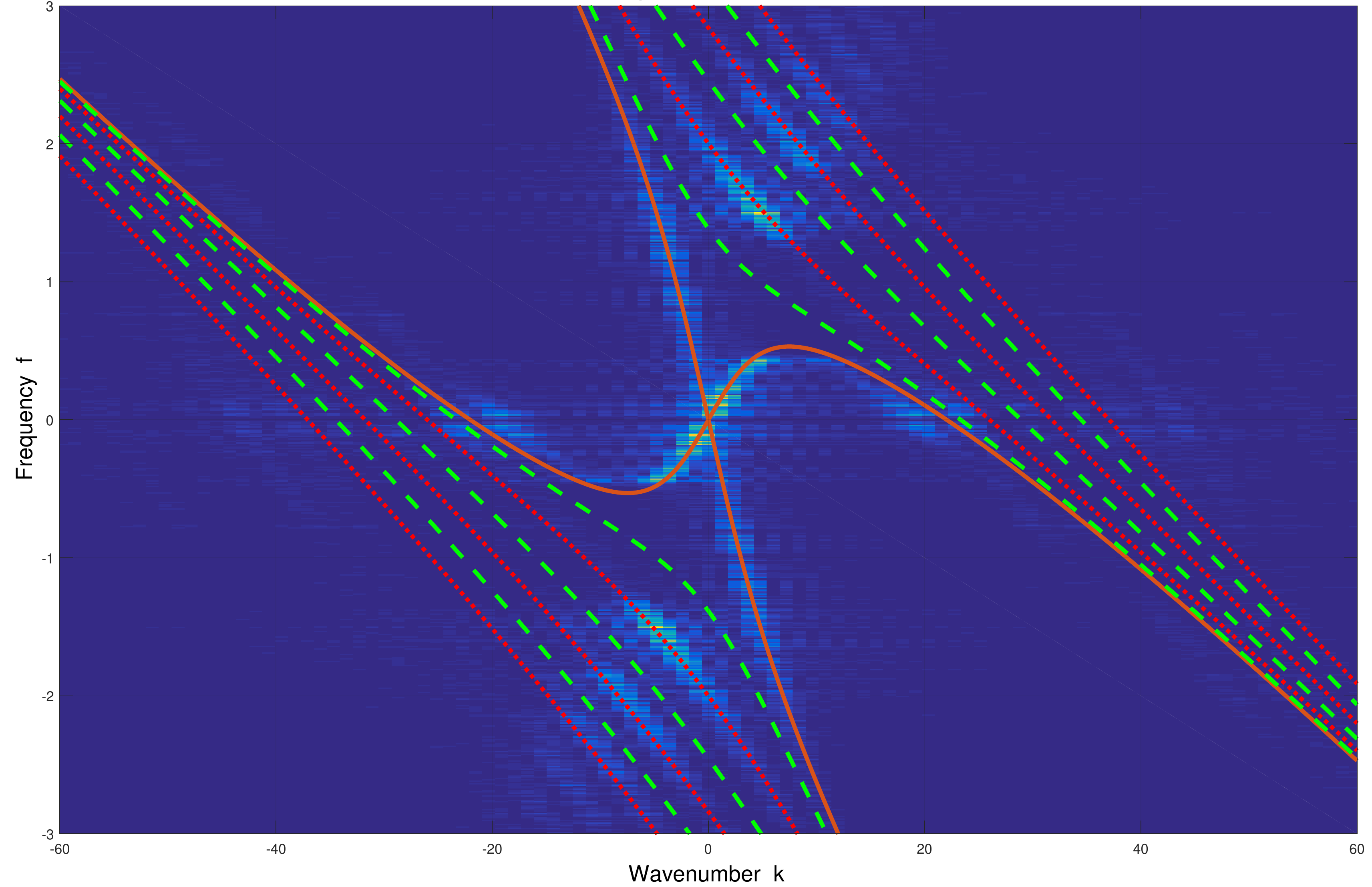}
\caption{(Left) Experimental setups for Analogue Gravity in Classical Hydrodynamics \cite{Rousseaux08, Rousseaux10, Weinfurtner11, Como11, Lawrence12, Vazquez12, Faltot13, Faltot14, Euve14, Euve15, Euve16a, Euve16b, Euve17, EuvePhD17, Coello17, Soto19, Euve20}: the water channel of the Institut Pprime with a laser sheet which illuminates the water column colored by a fluorescent dye ; the flow goes over an obstacle and waves may be stimulated at the other end of the channel by an oscillating weir. (Right) Dispersion relation of pure gravity waves $f=\omega/2\pi =({\bf U.k}\pm \sqrt{gk\tanh (kh)})/2\pi$ (measured in absence of an obstacle for a turbulent flow current) featuring both positive and negative energy branches with the transverse modes akin to massive particles in astrophysics \cite{Jannes11b} or jumping salmons exploring another dimension in real rivers (the even (red dots) and odd (green dots) modes are theoretically represented but only the even modes are measured with a central laser sheet) \cite{Weinfurtner11, Faltot13}. Parameters are: water depth $h=0.24m$, 3 cameras concatenated fields of view $L=2.80m$, acquisition frequency $f_a=30Hz$, recording time $t_r=120s$, 2048 images, flow rate per unit width $q=0.16m^2/s$, mean speed $U=q/h=0.668m/s$ (from the Master and PhD Theses of Pierre-Jean Faltot \cite {Faltot13} and L\'eo-Paul Euv\'e \cite{Euve14} at the Poitiers University under the supervision of Germain Rousseaux).}
\label{channel}
\end{figure}

\begin{figure}[!htbp]
\includegraphics[width=2.6in]{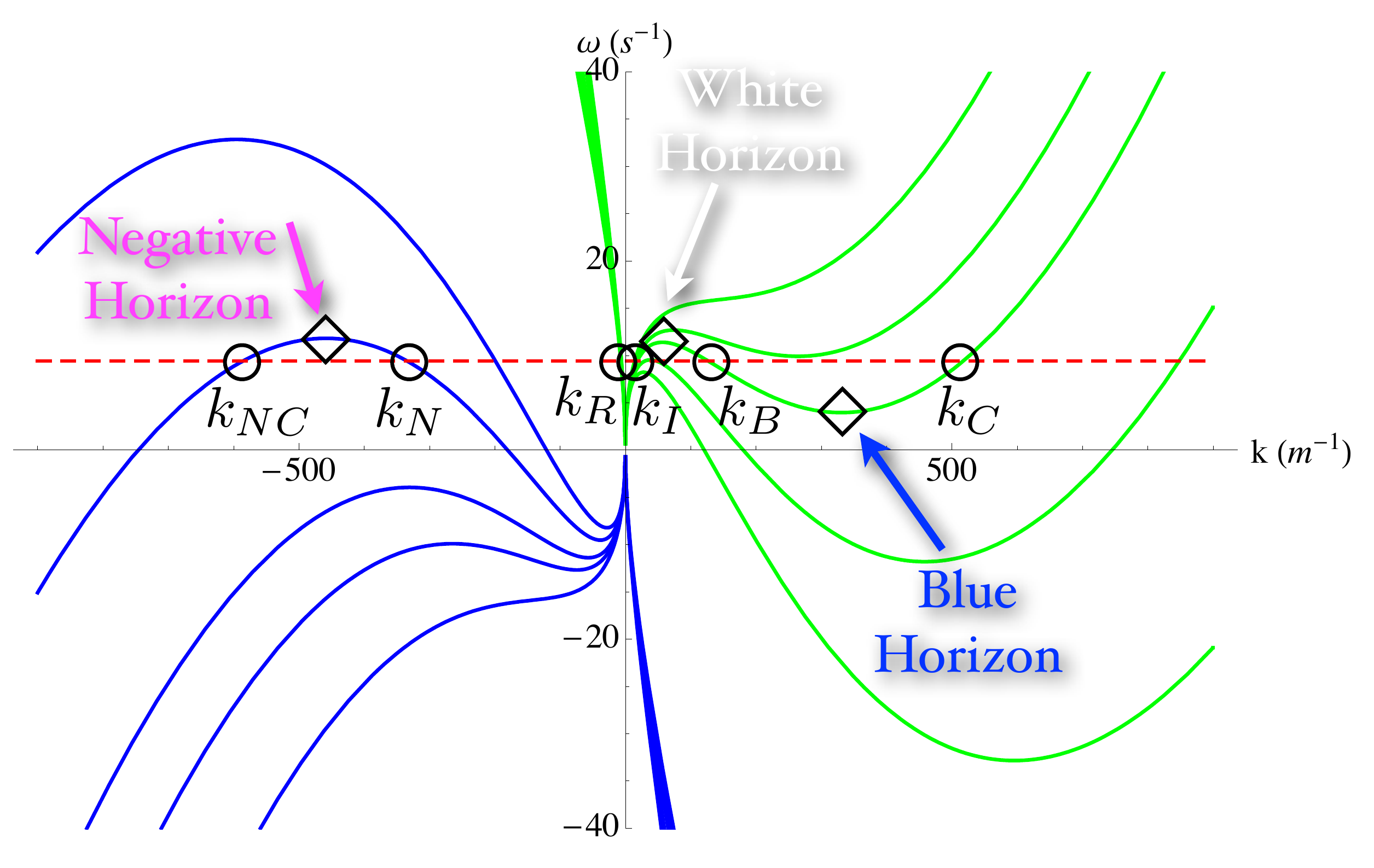}
\includegraphics[width=2.6in,height=1.7in]{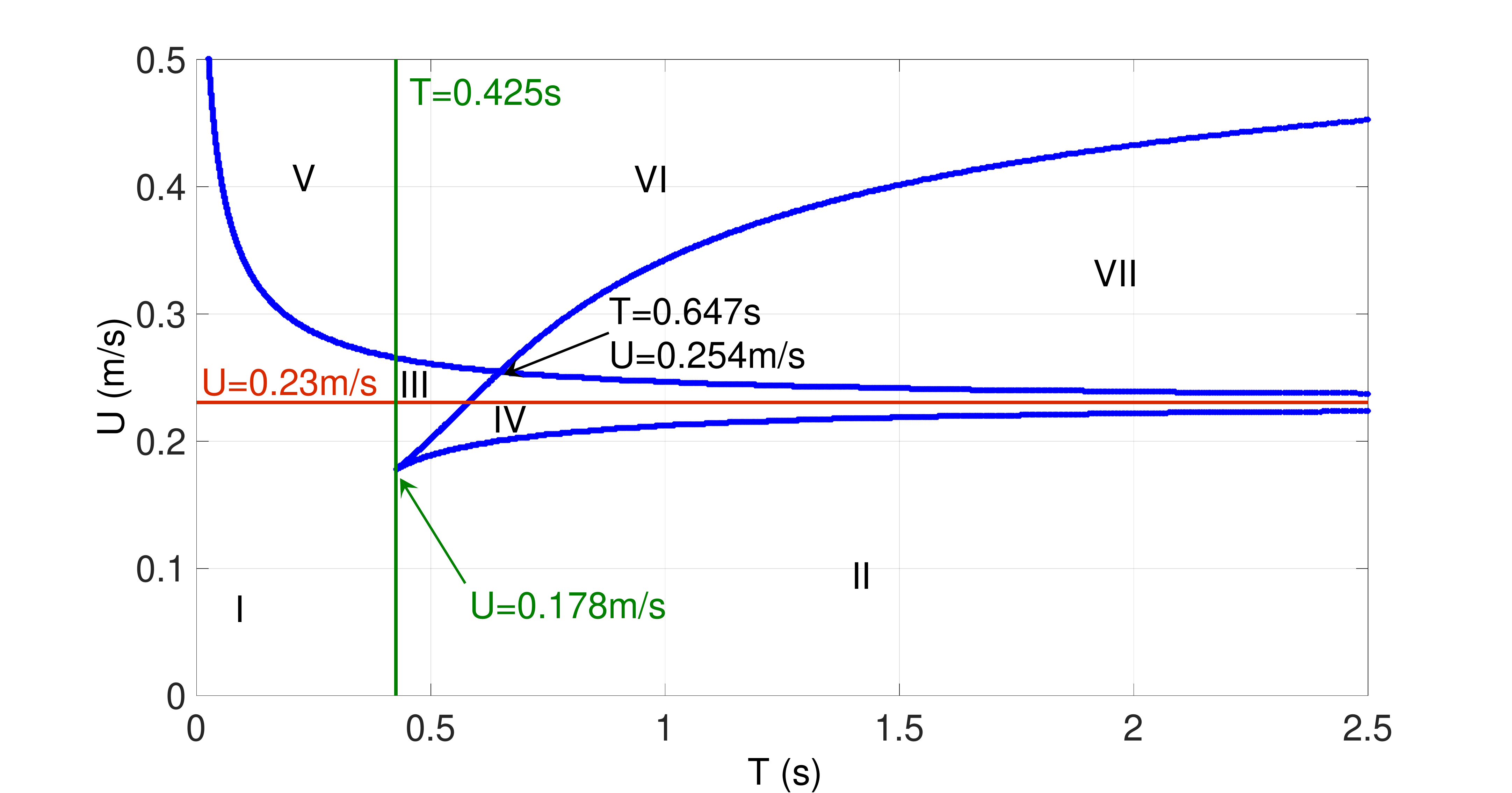}
\caption{(Left) Dispersion relation $\omega ={\bf U.k}\pm \sqrt{\left(gk+\frac{\gamma}{\rho}k^3\right)\tanh (kh)}$ in the laboratory frame of gravito-capillary waves in presence of a flow current with its two sub-branches (positive $+$ in green and negative $-$ in blue) and six solutions in the counter-current case for several speeds values \cite{Rousseaux10, Como11}. (Right) Theoretical wave phase diagram (computed for a water depth $h=0.05m$ \cite{Como11}): up to seven regimes are theoretically derived for the wave-blocking velocity $U$ of the counter-current as function of the period $T=2\pi/\omega$ of the incoming water waves \cite{Nardin09, Rousseaux10, Como11}. The water depth and capillary length play the role of Planckian scales (in Astrophysics, the Planck scale is given by $\sqrt{h_PG/c_L^3}$ where $h_P$ is Planck's constant, $c_L$ is the speed of light and $G$ is the gravitational constant) where Quantum Gravity effects would break Lorentz covariance in Astrophysics.}
\label{dispfull}
\end{figure}

When looking to steady open channel flows in Classical Hydrodynamics like in the experimental setups of the Figure \ref{channel} where a streamlined obstacle is placed in the flume to create a spatially-varying current \cite{Rousseaux08, Rousseaux10, Weinfurtner11, Como11, Lawrence12, Vazquez12, Faltot13, Faltot14, Euve14, Euve15, Euve16a, Euve16b, Euve17, EuvePhD17, Coello17, Soto19, Euve20}, the dispersion relation of gravito-capillary waves in the flowing fluid (say water) is $(\omega -{\bf U.k})^2=\left(gk+\frac{\gamma}{\rho}k^3\right)\tanh (kh)$, where $h$ is the mean water depth, $g$ the acceleration due to gravity, $\gamma$ the surface tension, $\rho$ the fluid density. The "acoustic" dispersion relation is recovered for the shallow water limit when $kh<<1$ for an infinitesimal wave camber $ka<<1$ (no harmonics generation or wave breaking \cite{Rousseaux08, Euve16a}) with $a$ the wave amplitude and for negligible surface tension effect $kl_c<<1$ where $l_c=\sqrt{\gamma /(\rho g)}$ is the capillary length \cite{Rousseaux10, Como11}. In absence of flow current $U=0$, there are only two solutions for k, describing waves of same wavelength but propagating in opposing directions. When $U \neq 0$ and taking into account the fluid capillarity, there are up to six solutions for the wavenumber $k$ (see Figures \ref{channel} and \ref{dispfull}). $k_I$, $k_B$, $k_C$ and $k_R$ stand for the incoming, blue-shifted, capillary and retrograde wave-numbers. Note the negative relative frequency for the negative gravity waves $k_N$ and negative capillary waves $k_{NC}$ \cite{Rousseaux10, Como11}. 

The noise due to the bulk turbulent flow (for which the Reynolds number $Re=Uh/\nu$ is large with $\nu$ the fluid viscosity) induces a wave turbulence noise on the free surface, which is analogous to the quantum noise in space-time: this noise is easily measured and leads to the dispersion relation of the system for flat analogue space-times (see Fig. \ref{channel} and \cite{Weinfurtner11, Faltot13, Faltot14, Euve16b, EuvePhD17}). Hence, one can look to the "spontaneous" amplification (non-stimulated by a wave-maker) by the flow velocity gradient at the horizon (the analogue of Hawking temperature) of the noise, which is what the analogue gravity community defined as the analogue Hawking Radiation. A quick remark on terminology: by "spontaneous", it is usually understood that spontaneous emission in free space depends upon vacuum fluctuations to be initiated and that the pairs are also entangled in quantum mechanics. The Universal Hawking radiation has nothing to do with entanglement: it may happen that the pairs are entangled if the system is quantum. Of course, in classical settings, we do not observe entanglement. In any cases, the free surface noise is amplified \`a la Hawking "spontaneously" from the point of view of the turbulent noise which plays the role of quantum noise with a double analogy \cite{Euve16b} (when the Strouhal or Unruh number $St=\omega / (dU/dx)_{U=c}$ is of order one \cite{Como11}). Since the terminology "spontaneous" is used by the quantum community to denote the quantum process, one may use the "unstimulated analogue amplification" process for the "spontaneous" amplification of the mechanical noise usually seen as a "classical stimulated amplification". Hence, the superfluid or quantum optical analogues are relevant only in the study of the additional effect of quantum entanglement \cite{Drori19, Nova19, Isoard20, Nova20}. The question of thermality of the spectrum has to do with the amplification process of the positive modes by the simultaneous generation of the negative modes and can be tested both in classical (to be done, see \cite{Euve14, Michel14, Euve16a, EuvePhD17, Michel18a}) and quantum experiments (to be done, see \cite{Nova19, Isoard20}) for transcritical flows without dispersive effects, zero modes for dispersive white hole flows, non-linear effects, dissipation effects, etc...

Long gravity waves are usually generated downstream of the bottom obstacle by a wave-maker and propagate upstream against the current \cite{Rousseaux08, Rousseaux10, Weinfurtner11}. They are possibly blocked on the lee side of the obstruction and mode-converted into a pair of shorter deep-water gravity waves (with or without wave blocking depending on the flow and wave parameters). The first of the pair $k_N$ has both group and phase velocities pointing downstream, whereas the second $k_B$ has the peculiar property of having opposite group and phase velocities \cite{Nardin09}. The blue-shifted modes $k_B$ are the classical analogues of the Hawking radiation when they are amplified by the presence of negative modes $k_N$. The latter are such that their frequencies in the co-moving frame of the water current are negative, hence their energies and their norms (the conserved quantity due to phase invariance of the wave equation or wave action in terms of water depths). We recall that the negative sign of the norm determines the anti-particle character in Quantum Field Theory \cite{Schutzhold02, Como11}. The reader will find an extensive discussion on the basics of Hydrodynamics for Analogue Gravity in the review chapter of the Como summer school (\cite{Como11}, Chapter 7). Three dispersive horizons are derived by cancelling the group velocity for white hole type flows (the incoming waves propagates on a counter current in the stimulated regime): the white hole horizon for the incoming waves which are transformed into blued shifted waves, the blue horizon for the blue-shifted waves which are converted into capillary waves and the negative horizon where negative modes become negative capillary modes \cite{Rousseaux10, Como11}. The blue and negative horizon merge for the long period limit which corresponds to the so-called Landau speed $U_L$ that controls the appearance of not only the blue and negative modes but also the zero-frequency solution known as the undulation (or 1D wake of the obstacle of Figure \ref{channel}). In the pure gravity limit \cite{Unruh08}, the Landau speed $U_L$ is null in deep water (blue shifted and negative modes are always present whatever the speed value \cite{Nardin09}) whereas the Landau speed is equal to the non-dispersive wave speed $c$ in the shallow water limit without surface tension where the group, phase and negative horizons merge altogether \cite{Como11}. However, the surface tension induces a threshold in speed for both the appearance of blue and negative modes (namely Hawking radiation) and the undulation \cite{Rousseaux10, Como11} for usual intermediate water depths in open channel flows. The latter case corresponds to the experimental implementations of the analogy: in practice the flow speed must be greater than $U_L=\sqrt{2}\left(\frac{\gamma g}{\rho}\right)^{1/4}=0.231m/s$ in the asymptotic regions far from the obstacle in order to have Hawking radiation reaching the asymptotic observer to the detriment of the undulation appearance which induces a "super-Hawking" effect due to its velocity oscillations in space with an additional amplification \cite{Euve16b} (as far as white hole type flows are concerned). In the wave phase diagram of the Figure \ref{dispfull}, the Hawking effect (namely the simultaneous conversion of blue and negative modes) can only occur in the region VI (with blocking by a white hole horizon) and VII (without blocking in absence of a white hole horizon). The region V does feature the negative modes albeit without a white hole horizon. Regions I, II and IV does not feature a white hole horizon contrary to region IV (the wormhole travel experiment is done in this region in absence of negative modes \cite{Euve17}).

\begin{figure}[!htbp]
\includegraphics[width=4.5in]{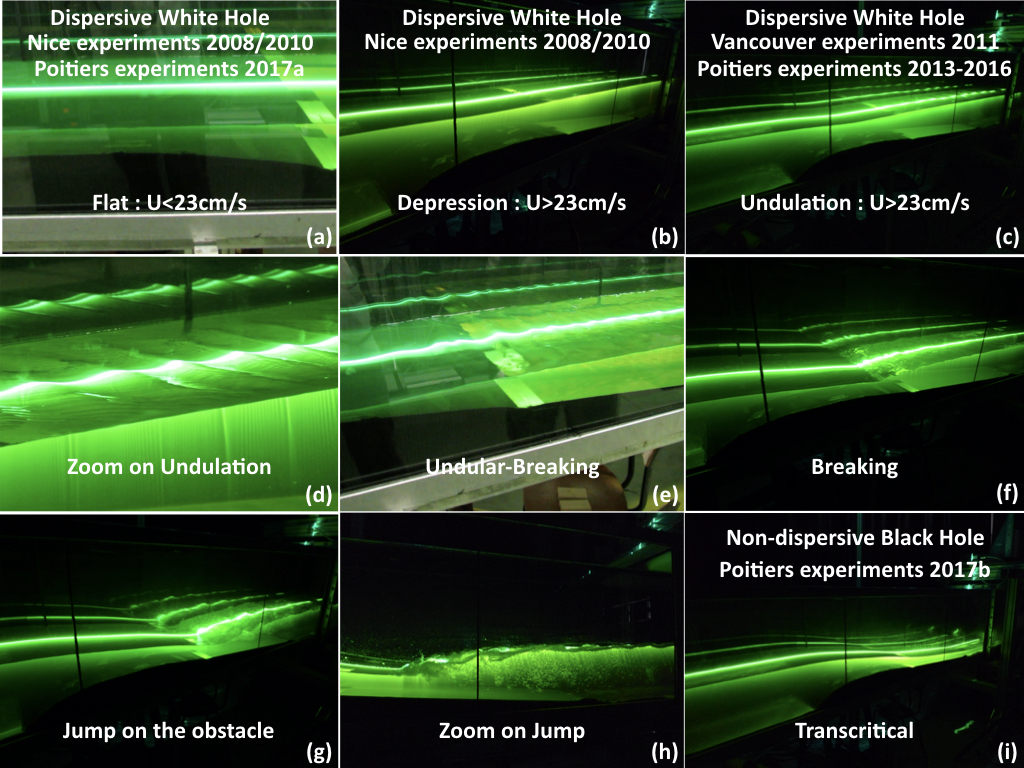}
\includegraphics[width=5.5in,height=2.5in]{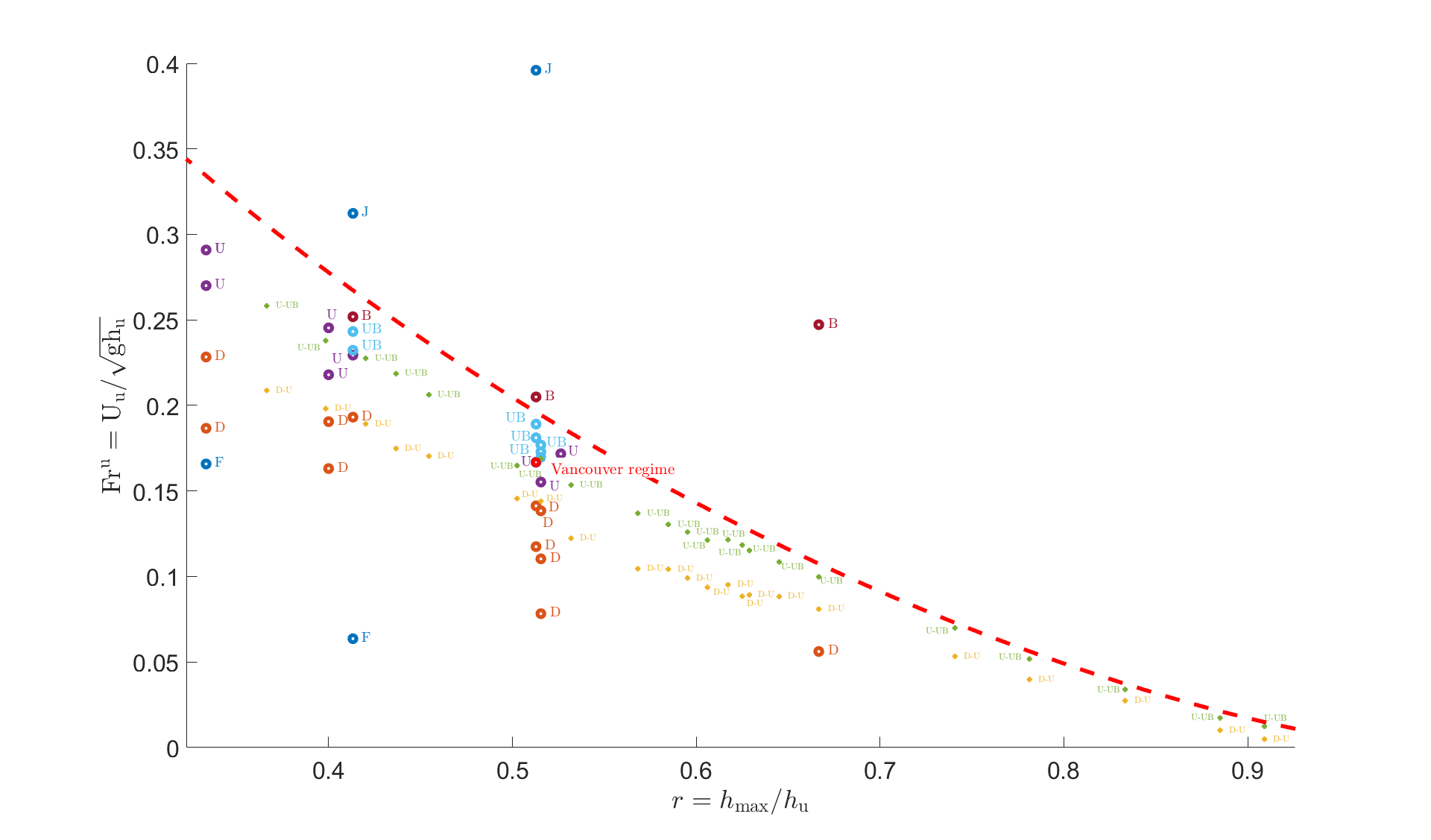}
\includegraphics[width=2.5in,height=2in]{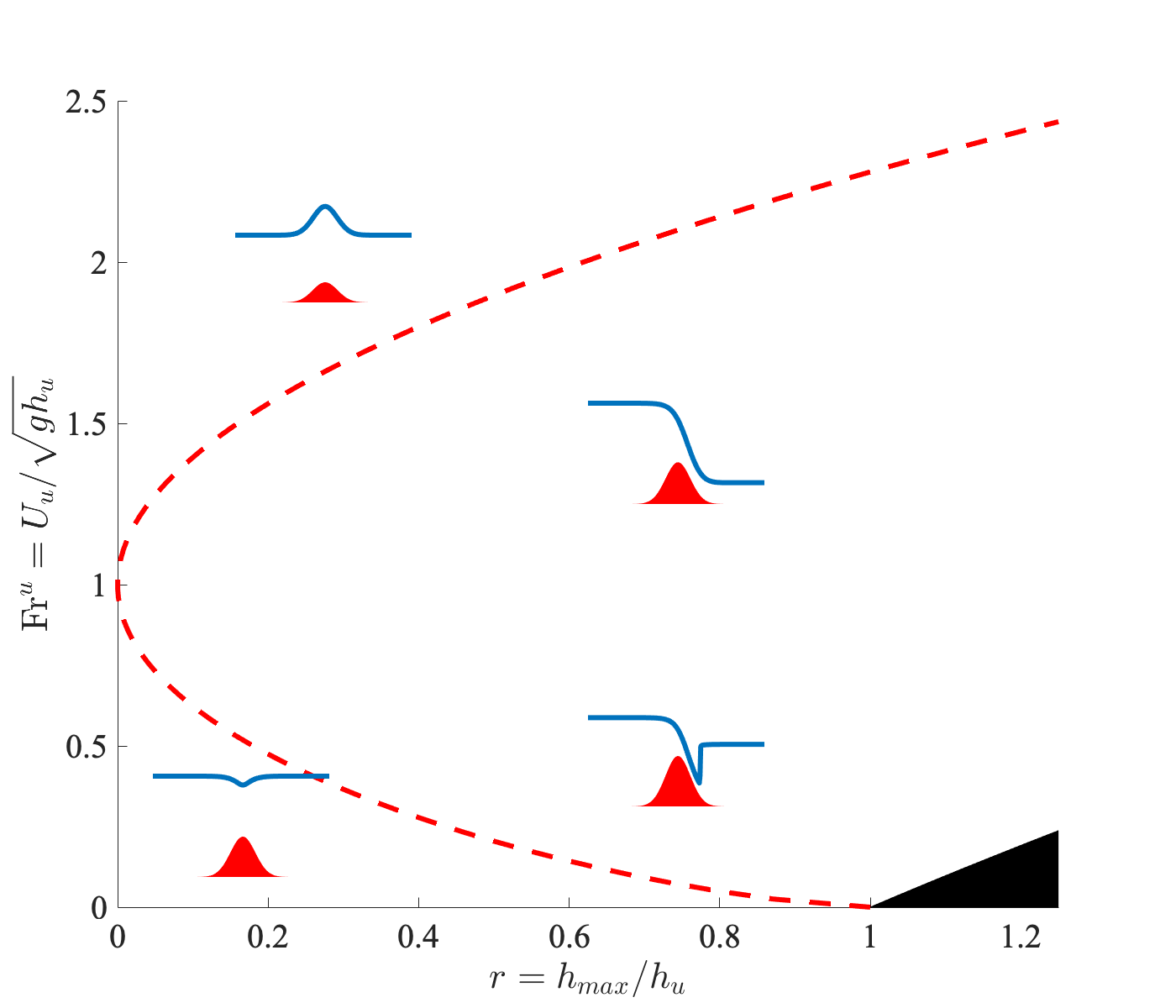}
\includegraphics[width=2.5in,height=2in]{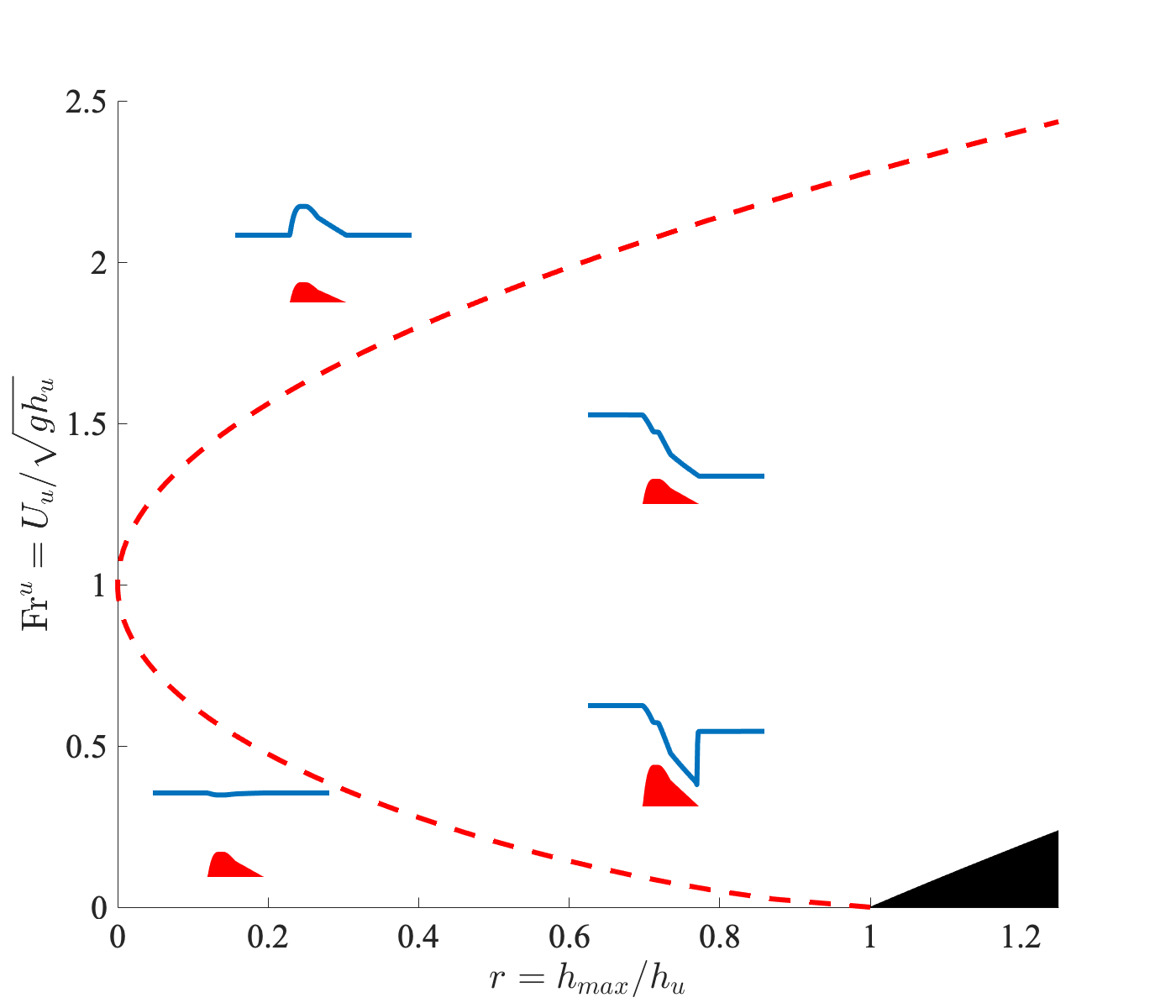}
\caption{ (Top) A bird eye view on the flow regimes over the Vancouver obstacle by increasing the flow rate at fixed initial water depth: visual observations with an Argon laser sheet and fluoresceine dye ; (Middle) Experimental flow phase diagram (from the Master and PhD Theses of Pierre-Jean Faltot \cite {Faltot13} and L\'eo-Paul Euv\'e \cite{Euve14} at the Poitiers University under the supervision of Germain Rousseaux) ; (Bottom) Numerical flow phase diagrams for a Gaussian obstacle and the Vancouver obstacle obtained by solving the Saint-Venant equations (by courtesy of Julien Dambrine, LMA, Poitiers University).}
\label{phase}
\end{figure}

In addition to the wave phase diagram computed from the dispersion relation (see Fig. \ref{dispfull}), a hydraulic phase diagram can help the experimentalists to predict the effective space-time on which the waves will propagate (see Fig. \ref{phase}). It is currently an open problem for free surface flows past disturbance despite its long history and attempts at classification of types of flows and free surface deformation \cite{Larsen66, Lawrence87, Chanson95, Lowery99, Vigie05, Binder19, Castro-Orgaz19}. In this work, we will not enter into a general discussion on this matter but we review the main type of experiments performed so far. We choose the Vancouver geometry for the obstacle \cite{Weinfurtner11} but our discussion can apply to any type of geometry (Nice 2008 \cite{Rousseaux08}, Nice 2010 \cite{Rousseaux10}, Orsay \cite{Euve16b} or Poitiers \cite{Euve20}).

Experimentally, the flow phase diagram is obtained by plotting the upstream Froude number $Fr_u=\frac{U}{\sqrt{gh_u}}$ as a function of the obstruction ratio $r=\frac{h_{max}}{h_u}$ where $h_{max}$ is the maximum height of the bottom obstacle and $h_u$ is the upstream asymptotic water depth (see Fig. \ref{phase}). We distinguish the following flow regimes with respect to the deformation of the free surface as observed with a laser sheet and fluorescent dye : F = Flat free surface ; D = free surface Depression ; D-U = border between free surface depression and non-breaking undulation ; U = (non-breaking) Undulation ; U-UB = border between non-breaking Undulation and Undular-Breaking zero mode or undulation ; UB = Undular-Breaking zero mode or undulation ; B = Breaking undulation ; J = (hydraulic) Jump. We do not specify if the hydraulic jump is present on the top of the obstacle or swept on the downstream side of the obstacle as well as its type (undular or breaking). The red dot corresponds to the flow regime of the Vancouver experiments \cite{Weinfurtner11, Como11, Lawrence12}. Numerically, the simplest of hydraulic models is the 1D Saint-Venant set of equations (mass conservation $\partial_th+\partial_x(hU)=0$ and momentum conservation $\partial_t(hU)+\partial_x(hU^2+gh^2/2)=-gh\partial_xb$ where $b(x)$ is the bottom obstruction) which captures (after an initial transient regime) the transition from sub-critical to super-critical stationary flows with the dimensionless number $Fr_u=U/\sqrt{gh_u}$, the global upstream Froude number assuming uniform plug flow for the velocity profile and no dispersion \cite{Lawrence12, Castro-Orgaz19}. Because the free surface varies past a disturbance like a bottom obstacle or a lateral constriction, only the local Froude number pilots the trans-critical transition $Fr(x)=U(x)/\sqrt{gh(x)}=1$ which is the analogue of a dispersion-less horizon in General Relativity. The red dotted lines in the Figures \ref{phase} correspond to the situation when the local Froude number reaches 1 above the obstacle and are given by a relationship between the obstruction ratio $r$ and the upstream Froude number $r=1+0.5Fr_u^2-3/2Fr_u^{2/3}$ \cite{Vigie05}.

The Saint-Venant shallow water equations do not take into account the effect of dispersion: the reader will look to the steady and unsteady solutions of the forced KdV equation which is able to tackle some dispersive properties like the undulation despite the lack of a clear comparison between experiments and numerical calculations \cite{Lowery99, Binder19}. Thanks to the combination of the wave and flow phase diagrams (see Figures \ref{dispfull} and \ref{phase}) and depending on the flow rate and the initial water depth, the experimentalists can design all sorts of flow configurations \cite{Larsen66, Lawrence87, Chanson95, Lowery99, Vigie05, Binder19, Castro-Orgaz19} reproducing non-dispersive hydraulic black holes \cite{Euve20}, dispersive hydraulic white holes with \cite{Weinfurtner11, Euve16b} or without \cite{Rousseaux08, Rousseaux10} undulation and dispersive hydraulic wormholes \cite{Euve17} (the region in-between both horizons). The next generation of analogue gravity experiments in Classical Hydrodynamics shall combine both phase and flow diagrams by taking into account hydraulic and dispersive effects in order to design engineered analogue space-times. The hydraulic jump is an interesting topic of interest since it is reminiscent of the central singularity of a black hole which is solved here by a time-dependent structure when it is of the breaking type or by a transcritical undulation when it is stationary. Of course, the original kinematic and dispersionless analogy of White-Unruh \cite{White73, Unruh81} is broken but one could maybe learn something from the way Classical Hydrodynamics avoids a singularity...

\section{Analogue Gravity in Thin Film Flows}

We present here a very short review as well as new results on thin film flows for Analogue Gravity and some guidelines for future works. Their major interest relies in different dispersive properties and scales with respect to open channel flows. They have also common features with Bose-Einstein condensates.

\subsection{The circular hydraulic jump}

Thin films flows as produced in the kitchen sink like a circular jump are very interesting systems in Classical Hydrodynamics for Analogue Gravity since they probe different dispersive properties compared to open channel flows. The circular hydraulic jump is a perfect system in order to study the influence of a spatial curvature of a horizon on incoming waves in addition to superluminal dispersive effects due to the smallness of the water height. In particular, in the shallow water approximation $kh<<1$ keeping the effect of capillarity, the dispersion relation is similar to the one of phonons in Bose-Einstein condensates \cite{Nova19, Isoard20, Nova20}. The dispersive corrections are superluminal for large wave-numbers in both cases (the phase speed increases with $k$) and the dispersive scales can be tuned by the experimentalists: the healing length $\xi$ in a BEC \cite{Garay00} or a combination of the capillary length and the water depth ($\sqrt{l_c^2-h^2/3}$) in Classical Hydrodynamics \cite{Como11}.

\begin{figure}[!h]
\includegraphics[width=3in]{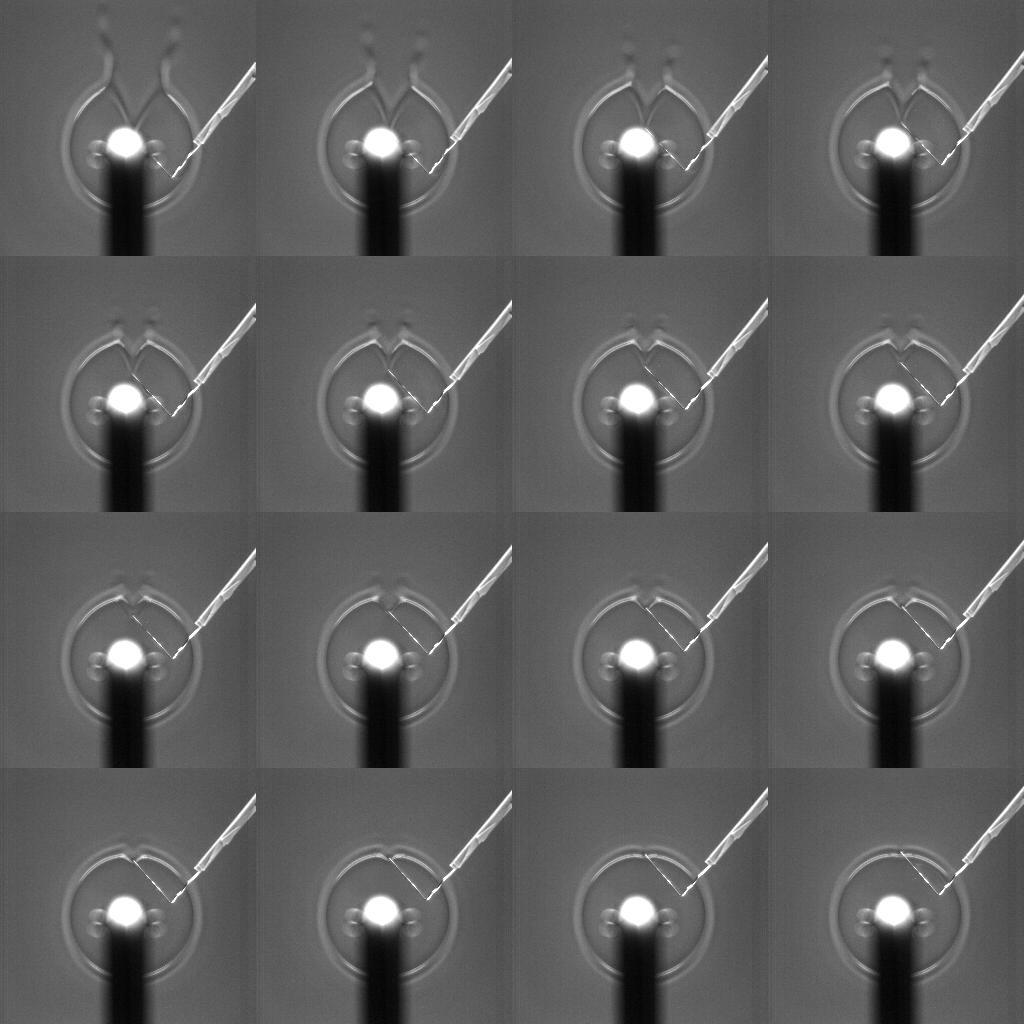}
\caption{Mach cones-like perturbations generated by a needle touching slightly the free surface inside the circular jump demonstrating a supercritical flow regime (the fluid is silicon oil) \cite{Jannes11a}.}
\label{montage}
\end{figure}

When a jet of tap water impacts a horizontal plate, a circular boundary between a supersonic region and a subsonic one is created and the water waves are either expulsed from the inner part or cannot enter from the outside when created by a perturbation (using a wave-maker for instance). Hence, a hydrodynamic white hole forms (the time reverse of a black hole that one cannot escape like a drain in a bathtub) that one cannot enter from outside \cite{Volovik05, Ray07, Jannes11a}. This phenomenon, easily observable, produces a rapid variation of the thickness of the liquid surface and of the current in space. Near the impact zone, there is a very thin layer of fluid of a few millimeters that thickens in a circular region located a few centimeters from the jet. By touching the free surface with a small needle in order not to perturb too much the flow, one observes inside the circular jump a Mach cone-like deformation which is controlled by a local Froude number $Fr(x)$ (to be defined according to the flow profile). This triangular wake is not only the demonstration of the acoustic metric but is also a signature of the super-critical character of the flow within the inner circle \cite{Jannes11a} (see Fig. \ref{montage}).

\begin{figure}[!h]
\includegraphics[width=2.5in]{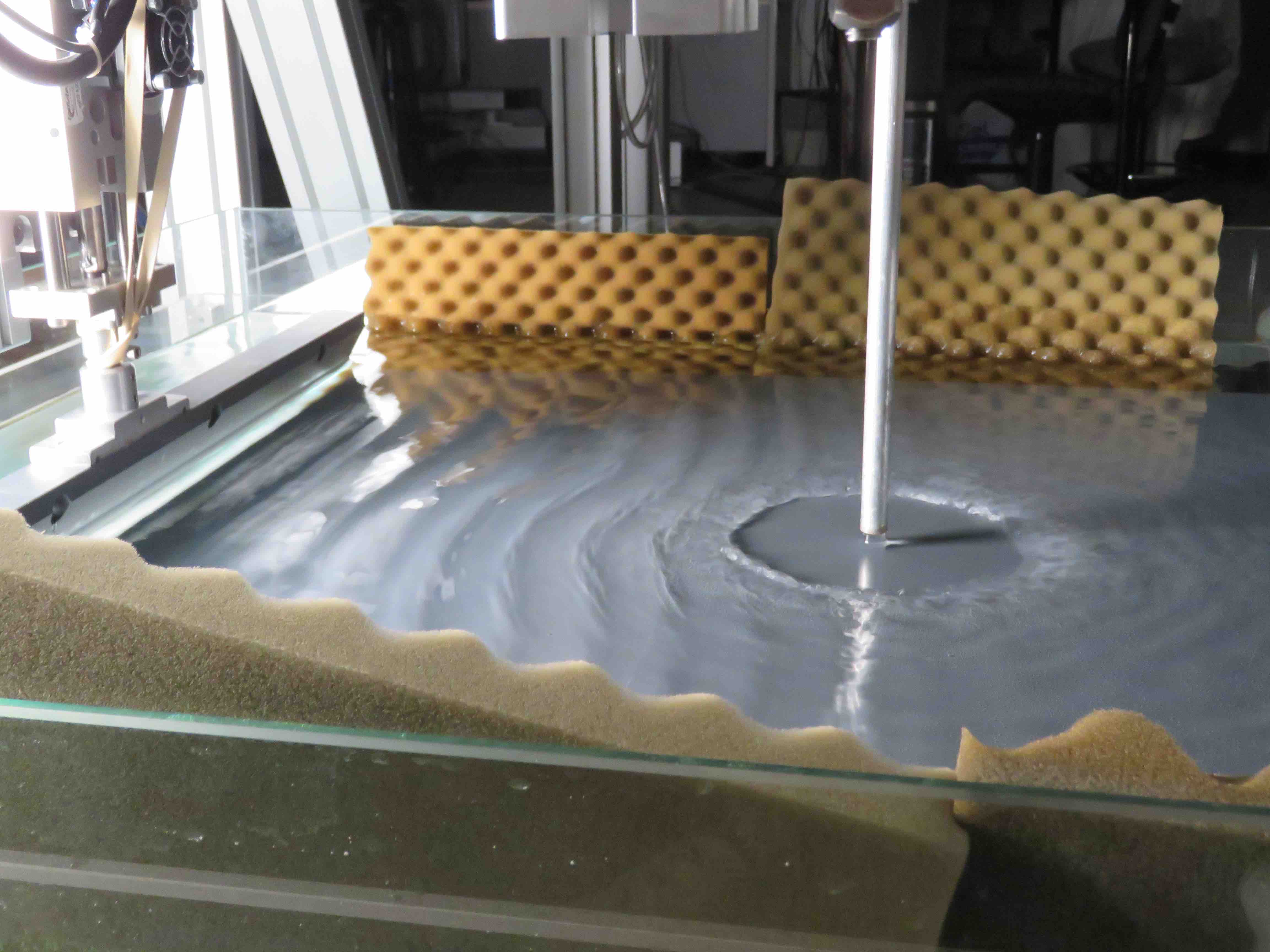}
\includegraphics[width=2.5in]{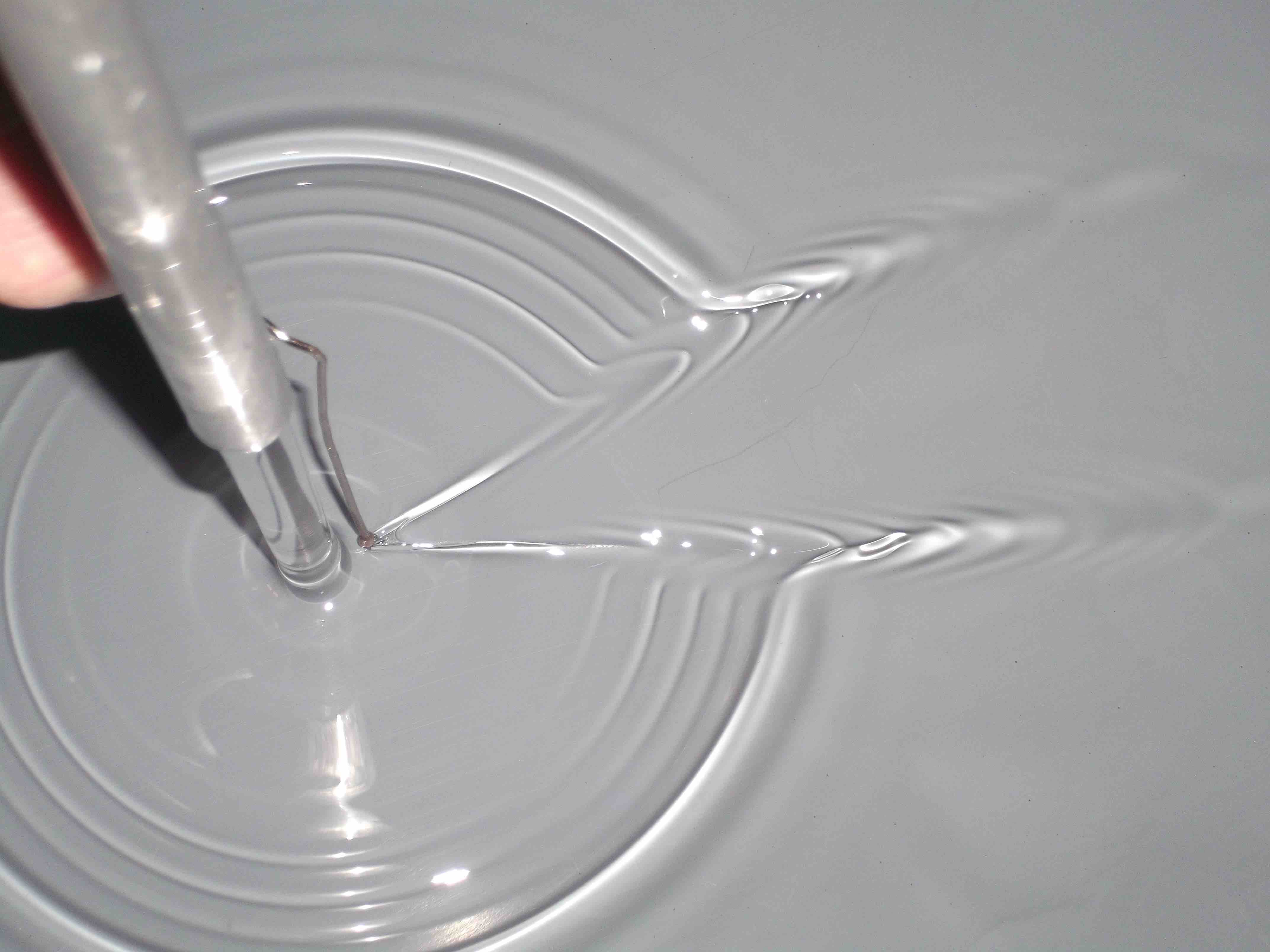}
\caption{(Left) The turbulent hydraulic white hole of the kitchen sink namely a circular hydraulic jump that incoming waves cannot enter from outside (the fluid is water). A wave-maker of the corner type is driven by a linear motor. The plane waves climb on an ascending slope before reaching a flat part where a vertical jet impacts and creates a circular jump. See the movie : \url{https://twitter.com/Ce_LZN/status/1166334141056606208} ; (Right) Mach cone inside the circular jump demonstrating a supercritical flow and the back-reaction on the outer jump induced by a needle plunging deeply (the fluid is silicon oil).}
\label{circularjump}
\end{figure}

We are currently studying its stimulated emission and scattering by sending water waves on/inside it in a manner similar to the recent studies on the scattering of a draining vortex by the team of Silke Weinfurtner in Nottingham \cite{Torres17, Torres18, Churilov19}. Finally, wave breaking will be reduced and viscosity can be changed by using different silicone oils in order to test several dispersion relations at the horizon tuned by surface tension and external imposed liquid height. The dispersive cut-off due to the viscosity will be particularly scrutinized \cite{Robertson18}. The quasi-normal modes encountered in General Relativity namely the way the circular jump comes back to equilibrium after a perturbation (a jump in the flow rate for example) will be also studied \cite{Torres18}. One of our goals is to characterize the diffractive-interference patterns of incoming waves on the white hole (see the Fig. \ref{circularjump}) by looking also to the effect of back-reaction when increasing the amplitude of waves \cite{Goodhew19}.

\subsection{Flowing soap films}

\begin{figure}[!h]
\includegraphics[width=2.5in, height=2.5in]{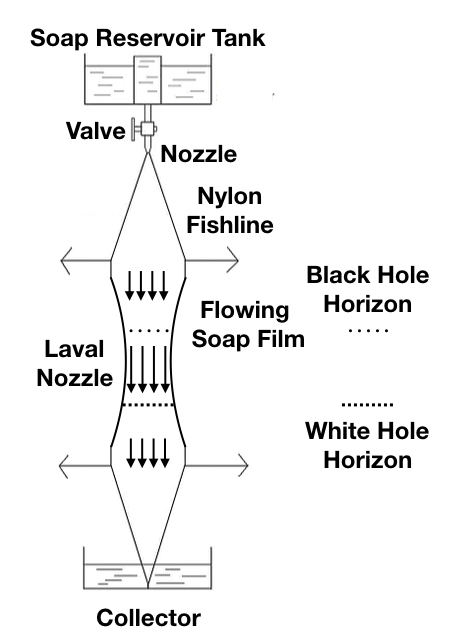}
\includegraphics[width=2.5in,height=2.5in]{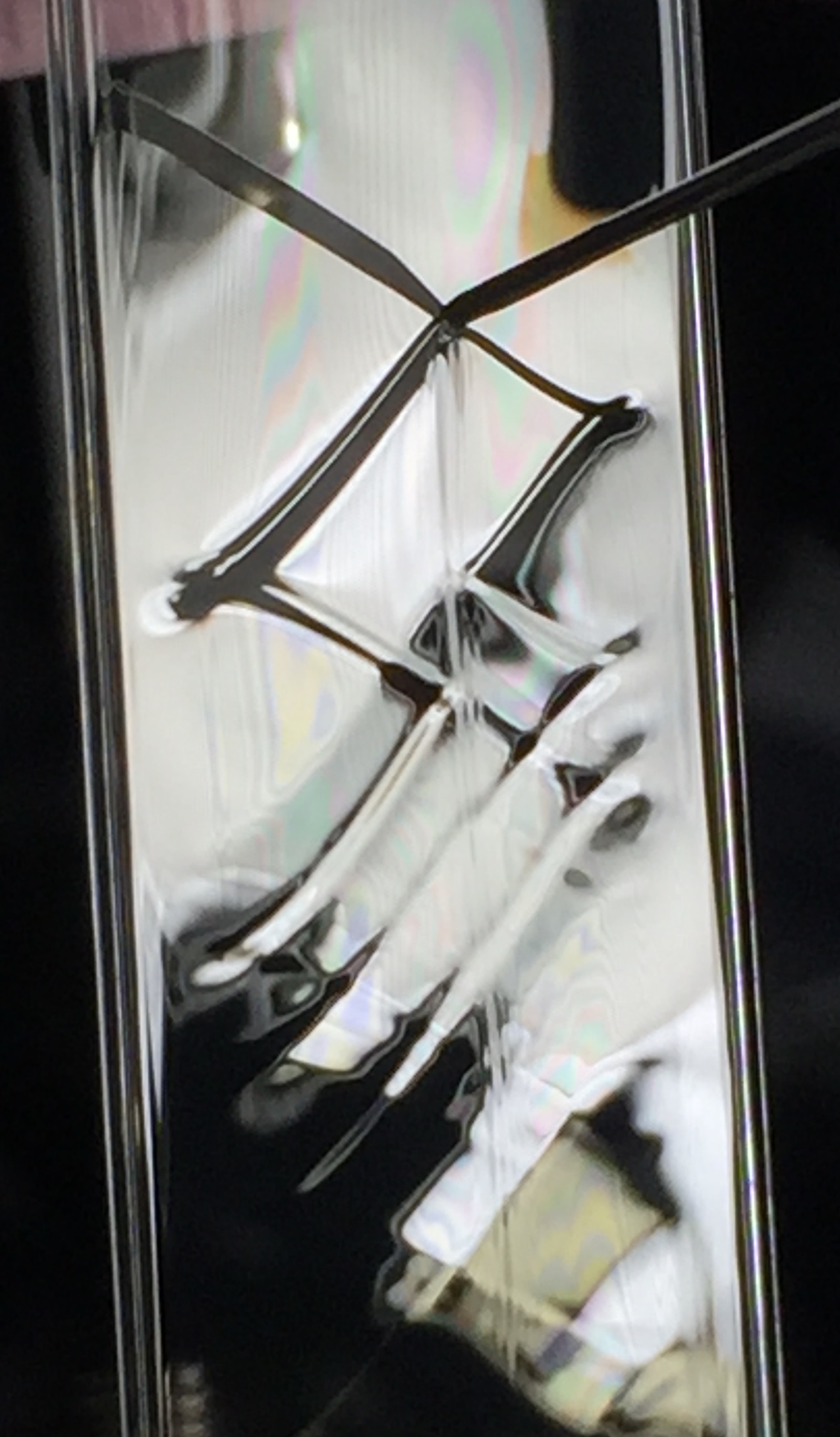}
\caption{(Left) A flowing soap film setup for an Analogue Gravity experiment. (Right) Mach cone-like deformation of a flowing soap film intercepted by a knife with its reflexions on the channel sides and its self-wrapping.}
\label{soaptunnel}
\end{figure}

The brand new system for analogue spacetimes that we are introducing consists in flowing soap films \cite{Kellay95, Kellay02, Meuel13, Kellay17} (or thin sheets of liquid \cite{Taylor59, Taylor74, Young81, Joosten85, Turner97}) whose width can be varied easily by the shape of Nylon fish-lines. The latter forms a narrowing geometry with a constriction leading to two trans-critical flow transitions. On the top of the soap films with either a vertical (see Fig. \ref{soaptunnel}) or horizontal geometry, capillary or elastic waves like bending or peristaltic waves \cite{Taylor59, Taylor74, Young81, Joosten85, Turner97} can propagate with a speed which depends on the surface tension/elasticity, the fluids densities of the liquid and the surrounding gas and the film thickness (the latter playing a role analogous to the water depth in an open channel flow). Hence, a soap film tunnel with non-dispersive horizons can be created easily and is certainly the cheaper experiment in Analogue Gravity after the circular jump.

\begin{figure}[!h]
\includegraphics[width=4.5in,height=3.5in]{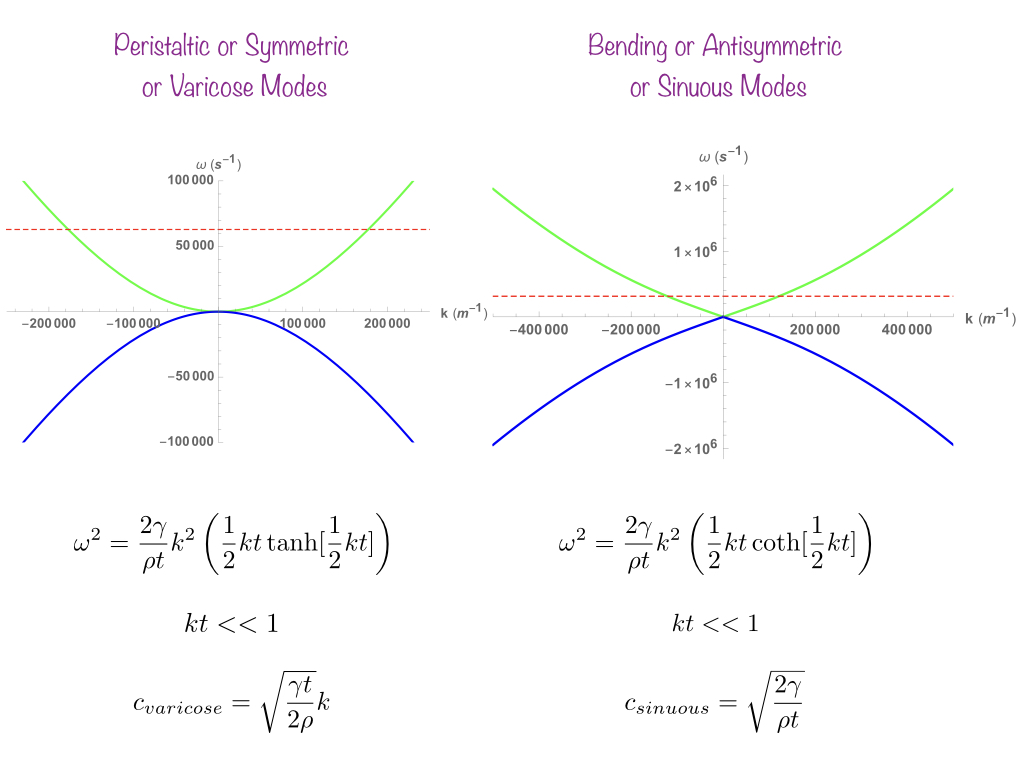}
\caption{The dispersion relations of hydrodynamic/elastic modes in soap films or thin sheets with the long wavelength limit of the corresponding wave speeds.}
\label{dispersionsoap}
\end{figure}

In 1959, G.I. Taylor studied the dynamics of thin sheets of liquids of thickness $t$. Two types of surface-tension-controlled waves (with their own dispersion relations, see Fig. \ref{dispersionsoap}) exist in these thin film system (in the limit of an inviscid ambient gas) in addition to an elastic mode for the first type: symmetrical waves (with respect to the sheet centerline corresponding to modulations of the median transverse position of the curtain) causing alternative thinning and thickening of the sheet (also called peristaltic or varicose modes) and anti-symmetrical waves causing sideways oscillating displacements of both sheet surfaces akin to a flapping flag (also called bending or sinuous modes) \cite{Taylor59, Taylor74, Turner97}.

\newpage

\begin{figure}[!h]
\includegraphics[width=4.5in,height=3.5in]{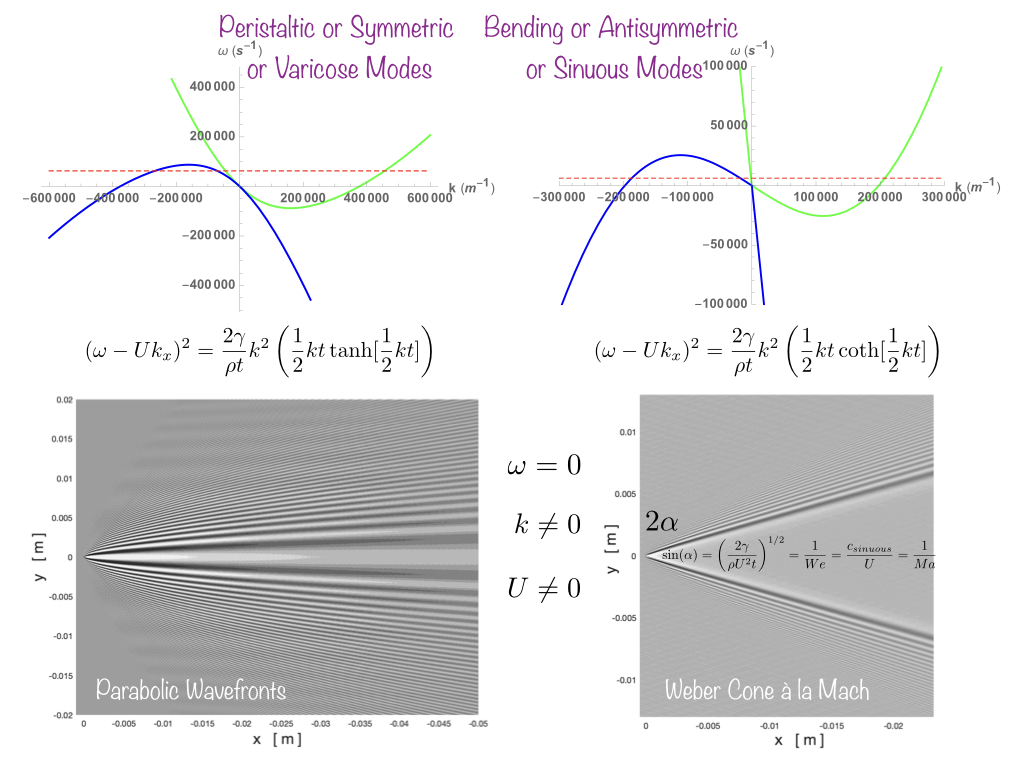}
\caption{(Top) Dispersion relations of mechanical waves of the capillary type on a flowing soap film (see Chapter 7 of \cite{Como11}) ; (Bottom) Corresponding Taylor's wakes \cite{Taylor59}.}
\label{wakesoap}
\end{figure}

The mathematical derivation of both dispersion relations by Taylor (see Fig. \ref{dispersionsoap}) is valid for small amplitudes and lead to the following expressions for the dispersion relations which are different because of different mass distributions in the film \cite{Taylor59, Young81, Joosten85}: $\omega ^2=\frac{2\gamma}{\rho t}k^2\left(\frac{1}{2}kt\tanh[\frac{1}{2}kt]\right)$ for the symmetric modes and $\omega ^2=\frac{2\gamma}{\rho t}k^2\left(\frac{1}{2}kt\coth[\frac{1}{2}kt]\right)$ for the antisymmetric modes. We include the effect of motion by the usual Doppler shift of the angular frequency $\omega '=\omega -Uk_x$ keeping the same wave-number $k'_x=k_x$ for 1D flows. The corresponding wakes are derived following the method described in the Chapter 7 of \cite{Como11} (see Fig. \ref{wakesoap}). A rigorous mathematical derivation would be welcome. Here we assume, as usual, that the lateral boundary layers are small with respect to the width of the channel. We neglect also vorticity and viscous effects: it was demonstrated that, in open channel flows, these complications modify quantitatively the picture but not qualitatively in most of the cases (see \cite{Como11}, Chapter 6 and \cite{Maissa16a, Maissa16b, Robertson18})...

In the presence of surfactants, the restoring force for the hydrodynamics anti-symmetric and symmetric modes is dominated by the surface tension $\gamma$ whereas Marangoni elasticity $E$ controls the elastic symmetric modes. The symmetrical waves form parabolic wavefronts (Chapter 7 of \cite{Como11} for a derivation) from a stationary disturbance (like a needle) and are dispersive with longer waves propagating more slowly for the entire wave-number range. The anti-symmetrical waves are non-dispersive only in the long wavelength limit (or for very thin sheets) with a non-dispersive celerity scaling with $t^{-1/2}$ (there is a degeneracy in the sense that both the symmetric capillary mode and the anti-symmetric elastic mode are possible with the same scaling for the speed). Taylor predicted that a small disturbance perturbing punctually a sheet expanding radially forms a cardioid wavefront \cite{Taylor59, Taylor74, Turner97}. 

A Mach cone-like deformation decorated by superluminal capillary waves is observed (see Fig. \ref{soaptunnel}) when a disturbance produced by a jet of air impinges on a sheet of water moving left to right for the anti-symmetric modes. Its angle is given by $\sin(\alpha)=\left(\frac{2\gamma}{\rho U^2t}\right)^{1/2}=\frac{1}{We}=\frac{c_{sinuous}}{U}=\frac{1}{Ma}$ where $W_e=\sqrt{\frac{\rho U^2t}{2\gamma}}$ is the Weber number comparing the effects of inertia and surface tension whereas the Mach number rules the transonic transition ($Ma=1$). The wavefronts were probed by a transmitted light in a Schlieren-like manner for the symmetrical waves and by reflexion pictures for anti-symmetrical waves \cite{Taylor59, Taylor74, Turner97}. Indeed, both elastic waves are present simultaneously so Taylor was compelled to observe them separately. The change of the sheet thickness induced by the symmetrical waves can reach as much as half its thickness and transmitted light is deflected by an angle which is higher where the thickness variation is stronger. On the contrary, the anti-symmetrical wave amplitudes can be many times bigger than the sheet thickness similarly to a stretched string with a radius much smaller than the wave amplitude. Hence, the angle of reflected light rays is larger than the one of transmitted rays whose angle of deflection is almost unchanged \cite{Taylor59, Taylor74, Turner97}. 

\begin{figure}[!h]
\includegraphics[width=4.5in,height=3.5in]{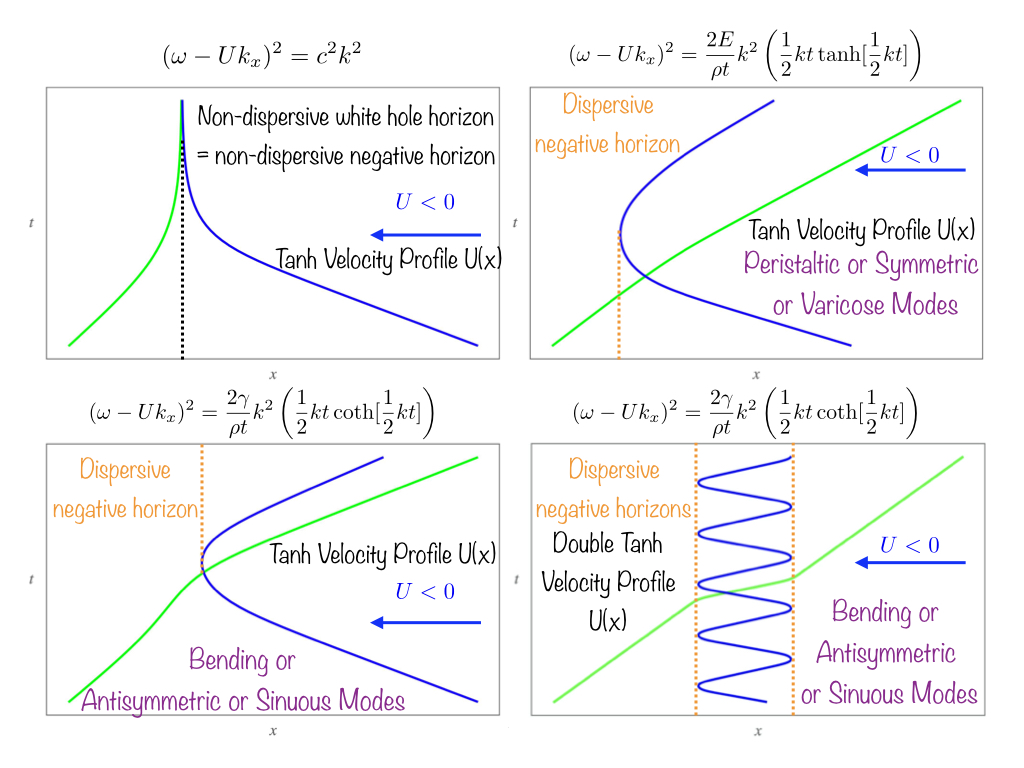}
\caption{Spacetime diagrams $x(space)$ vs $t(ime)$ for a white hole in a flowing soap film (see \cite{Rousseaux10, Peloquin16}). The flow is from right to left with negative speed $U<0$. The first three plots feature a simple $\tanh$ profile for the velocity field $U(x)$ whereas the fourth one has a double $\tanh$ profile allowing multiple reflexion of the negative modes.}
\label{xvstsoap}
\end{figure}

When $kt>>1$, one gets $c_{varicose}=c_{sinuous}$ and both mode propagates at the same speed. If $kt<<1$, the anti-symmetric mode is non-dispersive with a velocity given by $c_{sinuous}=\sqrt{\frac{2\gamma}{\rho t}}$ whereas the symmetric mode is dispersive with a velocity $c_{varicose}=\sqrt{\frac{\gamma t}{2\rho}}k$ that depends on the wave-number $k$. If one expands the dispersion relation in the anti-symmetric case, one recovers a dispersion relation similar to the one of a BEC with a superluminal behaviour and a quartic correction: the thickness $t$ of the soap film plays the role of the healing length $\xi$ in the condensate \cite{Garay00}. Since $c_{varicose}=\frac{kt}{2}c_{sinuous}$, the anti-symmetric mode is faster than the symmetric mode for very thin films and small deformation $ka<<1$ where $a$ is the amplitude of deformation. The assumption of small amplitudes of deformation was relaxed by Kinnersley in 1976 who derived exact non-linear solutions expressed as a function of both elliptic integrals and functions \cite{Kinnersley76}. When the thickness tends to infinity, the well-known non-linear capillary waves solutions in deep water of Crapper (1957) is recovered \cite{Crapper57}. The space-time behaviors are obtained by solving the rays equation with Mathematica as described in \cite{Rousseaux10, Peloquin16, Euve17} (see Fig. \ref{xvstsoap}). Indeed, the kinematics of short waves is ruled by Hamilton equations within the geometrical optics approximation: $\dot{x}=\partial \omega (x, k, t) / \partial k$ and $\dot{k}=-\partial \omega (x, k, t)/\partial x$ where the dot denotes the partial derivative with respect to time (the reader will not mix time with thickness in the space-time diagrams of Fig. \ref{xvstsoap}).

\begin{figure}[!h]
\includegraphics[width=4.5in,height=3.5in]{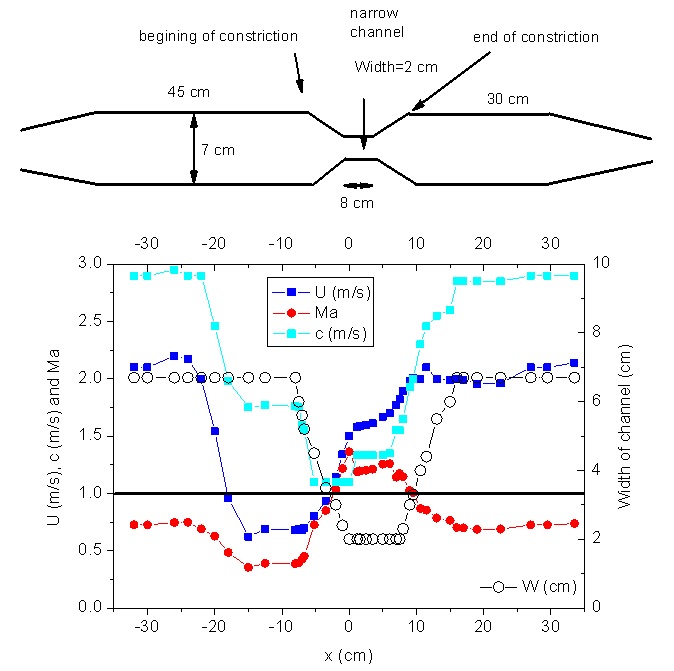}
\caption{A pair of black (on the left) and white (on the right) horizons ($Ma=1$) in a flowing soap film from left to right.}
\label{constrictionsoap}
\end{figure}

Preliminary experiments using soap film channels, such as that of Fig. \ref{soaptunnel} which displays the geometry of the channel and Fig. \ref{constrictionsoap} which displays the dimensions of the channel as well as measurements of the velocity profile along the channel, show that a transonic to supersonic transition as well as a supersonic to transonic transition regimes can be obtained (where the Mach number is $Ma=1$). Figure \ref{constrictionsoap} shows that the velocity of the flow can be smaller than the wave propagation speed in the areas of large channel width. However, and near the constriction, the flow velocity can exceed the wave velocity giving rise to a region where the wave speed lags behind the flow speed. The situation is reversed downstream as the channel width increases again. Typical channels are 2m in lengths with widths in the range of a few centimeters. The flow velocity is measured using Laser Doppler Velocimetry while the thickness of the film is measured (typical values are of order 10 micrometers) using a combination of velocity profiles and soap solution flux measurements. The velocity of the waves is deduced from measurements of the Mach cone angle and the velocity of the flow. Experiments are ongoing to examine how waves propagate in such a heterogeneous flow situation. Observations (not reported here) already indicate the existence of a standing wave pattern similar to the undulation in open channel flows that will be the subject of a future investigation... With flow and wave speeds of the order of magnitude of a few $m/s$, the surface gravity \cite{Visser98} of this type of flow is of the order of a few tens of $m/s^2$ (see Fig. \ref{constrictionsoap}). Hence, one expects a Hawking temperature of the order of $10pK$ similarly to the value obtained in open channel flows \cite{Euve16b} which is ten times smaller than the one for a BEC (around $100pK$ \cite{Nova19}).

\section{Conclusion}

\epigraph{{\it ``Now, as in a pun two truths lie hid under one expression, 
so in an analogy one truth is discovered under two expressions...Every question concerning analogies is therefore a question concerning the reciprocal of puns, and the solutions can be transposed by reciprocation.''}}{--- \textup{James Clerk Maxwell}, \textit{Are there real analogies in Nature?} (1856).}

What is Hawking radiation? It is an oxymoron: they are these sparks of cold light which would be consumed in the aether but that the cosmic maelstroms and fountains refract and magnify in the dark clarity of the Universe...\\

\vskip6pt

\enlargethispage{20pt}

{\bf Authors' Contributions:}

GR and his students designed and carried out the experiments on the open channel setups. HK carried out the experiments on the soap film flows. GR performed the data analysis on the soap film flows. GR and HK conceived and designed the soap film study, and GR drafted the manuscript with contributions by HK on the soap film part. Both authors read and approved the manuscript.\\

{\bf Funding:}

This work was supported by the French National Research Agency through the Grant No. ANR-15-CE30-0017-04 associated with the project HARALAB, by the University of Poitiers (ACI UP on Wave-Current Interactions 2013-2014), by the Interdisciplinary Mission of CNRS which funded the linear motor of the wave maker in 2013,  and by the University of Tours in a joint grant with the University of Poitiers (ARC Poitiers-Tours 2014--2015). It is currently funded by the project OFHYS of the CNRS 80 Prime initiative since 2019 and the Pprime ACI internal funding GrAnHysMice in 2020.\\

{\bf Acknowledgements:}

GR would like to thank his students Romain Piquet, Pierre-Jean Faltot, L\'eo-Paul Euv\'e, Johan Fourdrinoy and his colleague Julien Dambrine for their help in the preparation of this manuscript. Romain Bellanger, Laurent Dupuis and Jean-Marc Mougenot greatly contributed to the design, building and metrological developments of the experimental setups discussed in this work. HK would like to thank the IUF. We thank the organizers of the "The next generation of analogue gravity experiments" Royal Society meeting for allowing us to present this work.


\end{document}